\documentclass[12pt,preprint]{aastex6}
\usepackage{amsmath}
\usepackage{graphicx}
\usepackage{enumitem}
\usepackage{multirow}
\usepackage{color}

\shorttitle{Observational signatures of transverse MHD waves and associated dynamic instabilities}
\shortauthors{P. Antolin et al.}

\begin{document}

\title{Observational signatures of transverse MHD waves and associated dynamic instabilities in coronal flux tubes}

\author{P. Antolin\altaffilmark{1}, I. De Moortel\altaffilmark{1}, T. Van Doorsselaere\altaffilmark{2}, T. Yokoyama\altaffilmark{3}}
\affil{\altaffilmark{1}School of Mathematics and Statistics, University of St. Andrews, St. Andrews, Fife KY16 9SS, UK\\
\altaffilmark{2}Centre for mathematical Plasma Astrophysics, Mathematics Department, KU Leuven, Celestijnenlaan 200B bus 2400, B-3001 Leuven, Belgium\\
\altaffilmark{3}Department of Earth and Planetary Science, The University of Tokyo, Hongo, Bunkyo-ku, Tokyo 113-0033, Japan\\
}
\email{patrick.antolin@st-andrews.ac.uk}

\begin{abstract}

MHD waves permeate the solar atmosphere and constitute potential coronal heating agents. Yet, the waves detected so far may be but a small subset of the true existing wave power. Detection is limited  by instrumental constraints, but also by wave processes that localise the wave power in undetectable spatial scales. In this study we conduct 3D MHD simulations and forward modelling of standing transverse MHD waves in coronal loops with uniform and non-uniform temperature variation in the perpendicular cross-section. The observed signatures are largely dominated by the combination of the Kelvin-Helmholtz instability (KHI), resonant absorption and phase mixing. In the presence of a cross-loop temperature gradient we find that emission lines sensitive to the loop core catch different signatures than those more sensitive to the loop boundary and the surrounding corona, leading to an out-of-phase intensity modulation produced by the KHI mixing. Common signatures to all considered models include an intensity and loop width modulation at half the kink period, fine strand-like structure, a characteristic arrow-shaped structure in the Doppler maps, overall line broadening in time but particularly at the loop edges. For our model, most of these features can be captured with a spatial resolution of $0.33\arcsec$ and spectral resolution of 25~km~s$^{-1}$, although severe over-estimation of the line width is obtained. Resonant absorption leads to a significant decrease of the observed kinetic energy from Doppler motions over time, which is not recovered by a corresponding increase in the line width from phase mixing and the KHI motions. We estimate this hidden wave energy to be a factor of $5-10$ of the observed value.

\end{abstract}


\section{Introduction}

Research in the last decade has shown that transverse MHD waves permeate the solar atmosphere  \citep{Aschwanden_1999ApJ...520..880A,Nakariakov_1999Sci...285..862N, DeMoortel_Nakariakov_2012RSPTA.370.3193D,Tomczyk_2007Sci...317.1192T,Lin_2011SSRv..158..237L,Arregui_2012LRSP....9....2A,Anfinogentov_2015AA...583A.136A}. These MHD waves are of particular interest due to their ability to carry large amounts of energy, thus potentially contributing significantly to coronal heating \citep{Uchida_1974SoPh...35..451U,Wentzel_1974SoPh...39..129W,McIntosh_2011Natur.475..477M}. A peculiar characteristic of these waves, particularly those with high amplitudes, has been their very fast damping, and the leading theory explaining this behaviour is resonant absorption for the case of standing modes \citep{Goossens_2002AA...394L..39G,Goossens_2006RSPTA.364..433G,VanDoorsselaere_2004ApJ...606.1223V,Goossens_2011SSRv..158..289G}, and mode coupling for the case of propagating modes \citep{Allan_2000JGR...105..317A,Pascoe_2010ApJ...711..990P,DeMoortel_2016PPCF...58a4001D,Terradas_2010AA...524A..23T,Verth_2010ApJ...718L.102V}. 

Resonant absorption (and mode coupling), is an ideal process of energy transfer between different wave modes \citep{Ionson_1978ApJ...226..650I,Hollweg_1987ApJ...312..880H,Hollweg_1988JGR....93.5423H,Sakurai_1991SoPh..133..227S}, and has been shown to be very efficient and robust \citep{DeMoortel_Nakariakov_2012RSPTA.370.3193D,Pascoe_etal_2011ApJ...731...73P,Terradas_2008ApJ...679.1611T}. Resonant absorption predicts that in the presence of density inhomogeneity, the global kink mode can resonantly couple with Alfv\'en waves of azimuthal character. In the classical picture of coronal loops with a density gradient between the inside and the outside of the flux tube, the resonance layer is expected to exist at the boundaries (edges) of loops. The global kink mode, which consists of a purely transverse displacement of the loop, then behaves as local azimuthal Alfv\'en waves for most of the oscillation time. This means that most of the displacement (and thus the energy) from such waves is azimuthal and localised in the loop shell rather than transverse and global. 

Quantifying the amount of wave power in the solar atmosphere is essential for determining the role of waves in coronal heating. Based on the above, the ideal instruments to detect most of the wave power in Alfv\'enic modes are spectroscopic instruments. However, even when using such instruments, LOS effects are expected to be important since Doppler velocities (and therefore wave power) are significantly reduced due to the loss of spatial velocity coherency from phase mixing and the LOS integration across multiple waveguides \citep{DeMoortel_Pascoe_2012ApJ...746...31D}. In theory the remaining wave power should still be detectable, albeit with sufficiently high resolution, in the line broadening and also in the distribution of power in the frequency spectrum. \citet{McIntosh_DePontieu_2012ApJ...761..138M} analysed \textit{CoMP} spectroscopic measurements of the global corona, where clear periodic Doppler shifts are observed \citep{Tomczyk_2007Sci...317.1192T, VanDoorsselaere_2008ApJ...676L..73V}. The authors showed that the non-thermal line widths are correlated with the Doppler components, strongly suggesting an important wave contribution in the line widths. Supporting this conclusion, also with \textit{CoMP} data, \citet{Morton_2016ApJ...828...89M} have shown that there is a power spectrum enhancement at 3~mHz everywhere in the corona, suggesting a p-mode origin for the kink waves \citep{Cally_2016GMS...216..489C,Khomenko_Collados_2015LRSP...12....6K}.

The wave energy content hidden in the line broadening of coronal lines is still unclear. Through Monte Carlo simulations and comparison with the \textit{CoMP} observations, \citet{McIntosh_DePontieu_2012ApJ...761..138M} estimate the wave amplitudes to be between $25$ and $56~$km~s$^{-1}$. Other observational reports place the non-thermal component peak at about $15-25~$km~s$^{-1}$ (with a long tail up to $40~$km~s$^{-1}$ or so) in coronal lines \citep{Hara_Ichimoto_1999ApJ...513..969H, Doschek_etal_2007ApJ...667L.109D, Hara_2008ApJ...678L..67H}, and with a tendency to larger values for LOS transverse to the loop plane, supporting an Alfv\'enic origin. Observations at higher resolution in transition region lines \citep{DePontieu_2015ApJ...799L..12D} or prominence observations in chromospheric lines show similar values, suggesting that non-thermal velocities and spatial resolution are uncorrelated to some extent \citep{Parenti_2007AA...469.1109P,Parenti_2015ASSL..415...61P}. Such results can be interpreted as evidence for Alfv\'enic turbulence, a scenario also supported by numerical modelling \citep[e.g.][]{Asgari-Targhi_2014ApJ...786...28A,Woolsey_2015ApJ...811..136W,Cranmer_2015ApJ...812...71C}. This scenario is further supported by power spectrum analysis of Doppler shift oscillations from \textit{CoMP}, performed by \citet{DeMoortel_2014ApJ...782L..34D}, in which a shift of the power is observed towards high frequencies at the loop apex and is interpreted as a possible signature of the expected turbulence cascade.
 
Besides the determination of wave energy, the second biggest obstacle for Alfv\'{e}nic wave heating of the corona is determining an efficient dissipation mechanism. The generation of turbulence from waves is considered as a possible solution to this problem. Spatial inhomogeneities, wave-to-wave interaction and, particularly, instabilities, can be major sources for turbulence. Alfv\'enic waves such as kink waves can become unstable due to shear motions (Kelvin-Helmholtz instability - KHI) at the boundary (shell) of flux tubes \citep{Terradas_2008ApJ...687L.115T}. This instability has been obtained in photospheric, chromospheric and coronal conditions \citep{Karpen_1993ApJ...403..769K,Ofman_1994GeoRL..21.2259O,Poedts_1997SoPh..172...45P, Ziegler_1997AA...327..854Z}. In coronal loops, this instability can be obtained even for small amplitude ($\lesssim$3~km~s$^{-1}$) standing kink modes \citep{Antolin_2014ApJ...787L..22A}. The ease with which the KHI can set in has been recently confirmed by \citet{Zaqarashvili_2015ApJ...813..123Z}, who consider straight, twisted and rotating magnetic structures. For the latter the twist plays an inhibiting effect on the instability, which, however, can occur when the kinetic energy of rotation is more than the magnetic energy of the twist. This has been recently confirmed numerically by \citet{Murawski_2016MNRAS.459.2566M}. 

In the formation of turbulence from KHI associated with transverse MHD waves a myriad of vortices and current sheets are formed along the flux tube, which may contribute to the heating through viscous and resistive dissipation, but also through (component) magnetic reconnection, as suggested in \citet{Antolin_2014ApJ...787L..22A}. We refer to such KHI vortices as Transverse Wave Induced Kelvin-Helmholtz vortices, or TWIKH rolls. Through mixing and heating, combined with LOS effects, TWIKH rolls lead to strand-like structure in intensity images in coronal loops \citep{Antolin_2014ApJ...787L..22A, Antolin_2016ApJ...830L..22A}, and thread-like structure in prominences \citep{Antolin_2015ApJ...809...72A}. The KHI associated with transverse MHD waves thus can not only have important consequences for the thermodynamic evolution of loops but also for their morphology. Not only does this model show that strands are not a unique feature of reconnection \citep{Klimchuk_2006SoPh..234...41K}, but also that such structures cannot be independent. Indeed, as recently shown by \citet{Magyar_2016ApJ...823...82M}, an initially multi-stranded loop would rapidly lose such state due to a transverse MHD wave and, more precisely, due to the KHI and phase mixing that is produced. 

The KHI combined with resonant absorption and phase mixing in a coronal loop has important observational consequences that can explain several commonly detected features in observations. In \citet{Antolin_2015ApJ...809...72A} we characterised a set of imaging and spectroscopic observables for such waves travelling in prominences, providing an explanation for the observed phase difference between the line-of-sight (LOS) velocity signal and the transverse (in the plane-of-the-sky - POS motion) observed by \textit{Hinode}/SOT and \textit{IRIS} \citep{Okamoto_2015ApJ...809...71O} and, potentially the thread-like structure that is usually seen accompanying transverse motions of prominence threads \citep{Okamoto_2016ApJ...831..126O}. Other predicted observables from the combined KHI and resonant absorption mechanisms were the thinning and fading in cool lines, accompanied with broadening and intensity enhancement in hot lines, providing further explanation for observed prominence threads disappearing in the \ion{Ca}{2}~H line at roughly $10^{4}$~K followed by appearance in the hotter \ion{Si}{4} line at $10^{5}$~K. 

Recently, in \citet{Antolin_2016ApJ...830L..22A} we have shown that for imaging instruments with a spatial resolution such as \textit{SDO/}AIA, TWIKH rolls can lead to an apparent decay-less oscillation for low amplitude kink modes, due to the combined effect of periodic brightening from the vortices and their coherent motion. This effect, combined with the intensity dimming or enhancement produced from the KHI mixing, provides a possible physical explanation for observed decay-less oscillations with AIA \citep{Nistico_2013AA...552A..57N,Anfinogentov_2013AA...560A.107A}, a phenomenon that seems to be ubiquitous in active regions \citep{Anfinogentov_2015AA...583A.136A}. Furthermore, we have shown that when a non-uniform temperature variation exists across the loop, different emission lines that are sensitive to different parts of the loop can catch the effect from the TWIKH rolls with varying degree, leading to a temperature (and thereby density) dependent period from the effects of phase mixing. Another explanation for the decay-less effect has also been proposed by \citet{Nakariakov_2016AA...591L...5N}, in terms of self-oscillations at the natural loop frequency produced by quasi-steady flows.

In this paper we examine the observable quantities from numerical simulations of standing transverse MHD waves in coronal loops, including both linear and non-linear effects such as the KHI, with forward modelling targeting both imaging and spectroscopic instruments. We focus in particular on determining the instrumental requirements to detect the mechanisms at play, for comparison with future coronal observations. The paper is organised as follows. In section~\ref{model} we explain the numerical setup and the forward modelling. In section~\ref{results} we analyse the results from the numerical model, which are then forward modelled and analysed in section~\ref{forwardresults}. In section~\ref{prediction} further forward modelling is performed targeting spatial and spectral resolution of current instruments to test the resilience of our results during observations. We discuss the findings and conclude in section~\ref{discussion}.

\section{Numerical Model}\label{model}

Coronal loops are observed to be very dynamic, continuously appearing and disappearing across temperature passbands in imaging instruments such as AIA. Due to the longer cooling timescales with respect to the heating timescales it is expected and observed that loops are usually in a state of cooling \citep{Viall_2012ApJ...753...35V}. Due to their higher densities with respect to the ambient corona (main reason why we actually observe them as distinctive features in AIA passbands), the plasma inside will be cooling faster than the ambient plasma, a scenario which leads to different temperatures with the surrounding. For this reason in this work we have considered 2 different initial atmospheres for our loop, one presenting a density and temperature contrast with the ambient corona but with uniform magnetic field throughout, and another with density and magnetic field contrast with the ambient corona but with uniform temperature throughout. In both cases we have hydrostatic pressure balance initially. 

\subsection{Initial setup}

\begin{figure}
\begin{center}
$\begin{array}{c}
\includegraphics[scale=0.3]{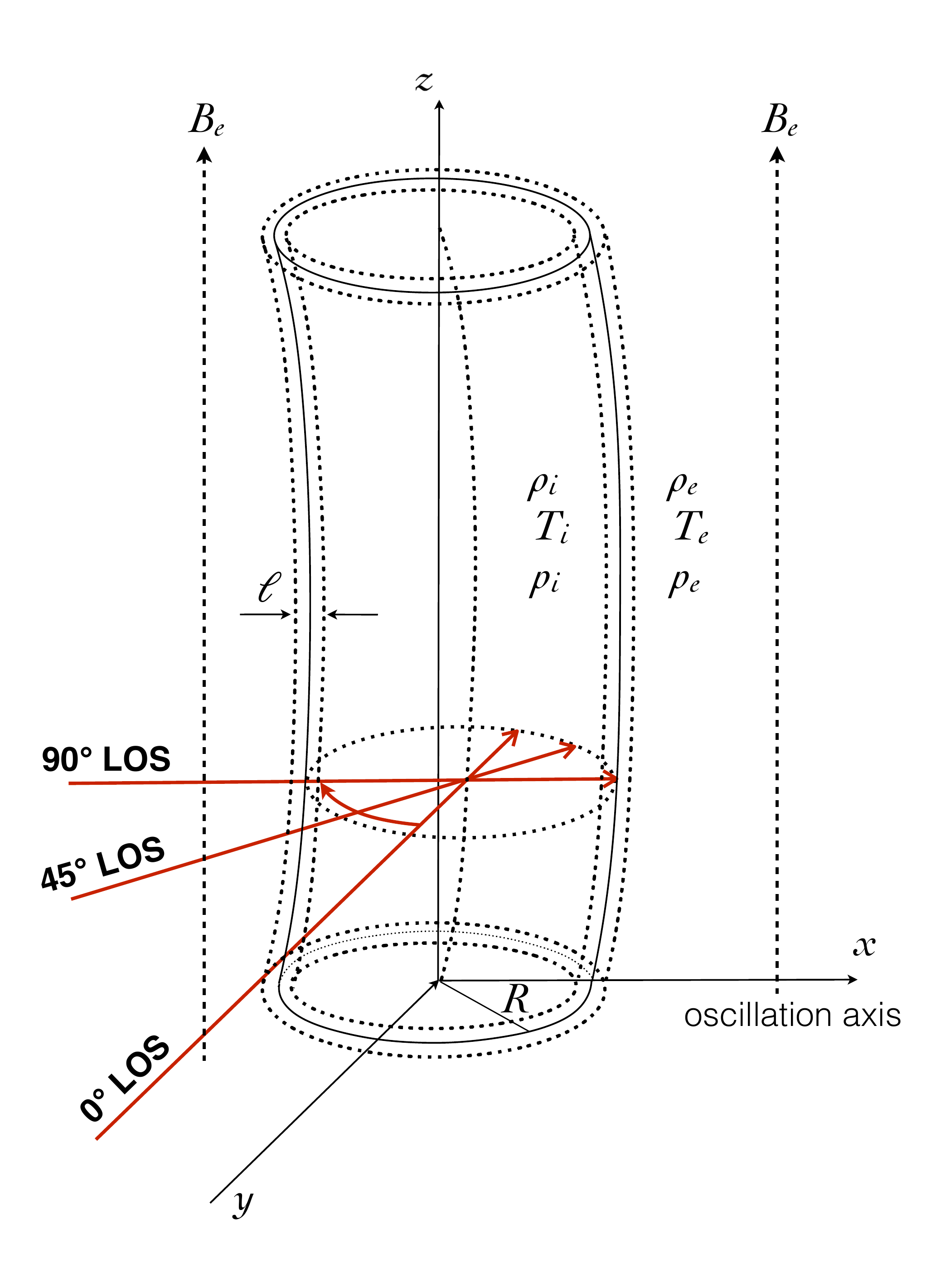}
\end{array}$
\caption{Sketch of the our loop with radius $R$ undergoing a fundamental kink mode perturbation along $x$. The quantities $\rho, T, p$ and $B$ denote the density, temperature, pressure and longitudinal magnetic field, respectively. The subindexes $i$ and $e$ denote the internal (within the loop) and external values (ambient corona). The LOS angles $0^{\circ}, 45^{\circ}, 90^{\circ}$ for the forward modelling are shown as red directional rays.
\label{fig1}}
\end{center}
\end{figure}

\begin{figure}
\begin{center}
$\begin{array}{cc}
\includegraphics[scale=0.35]{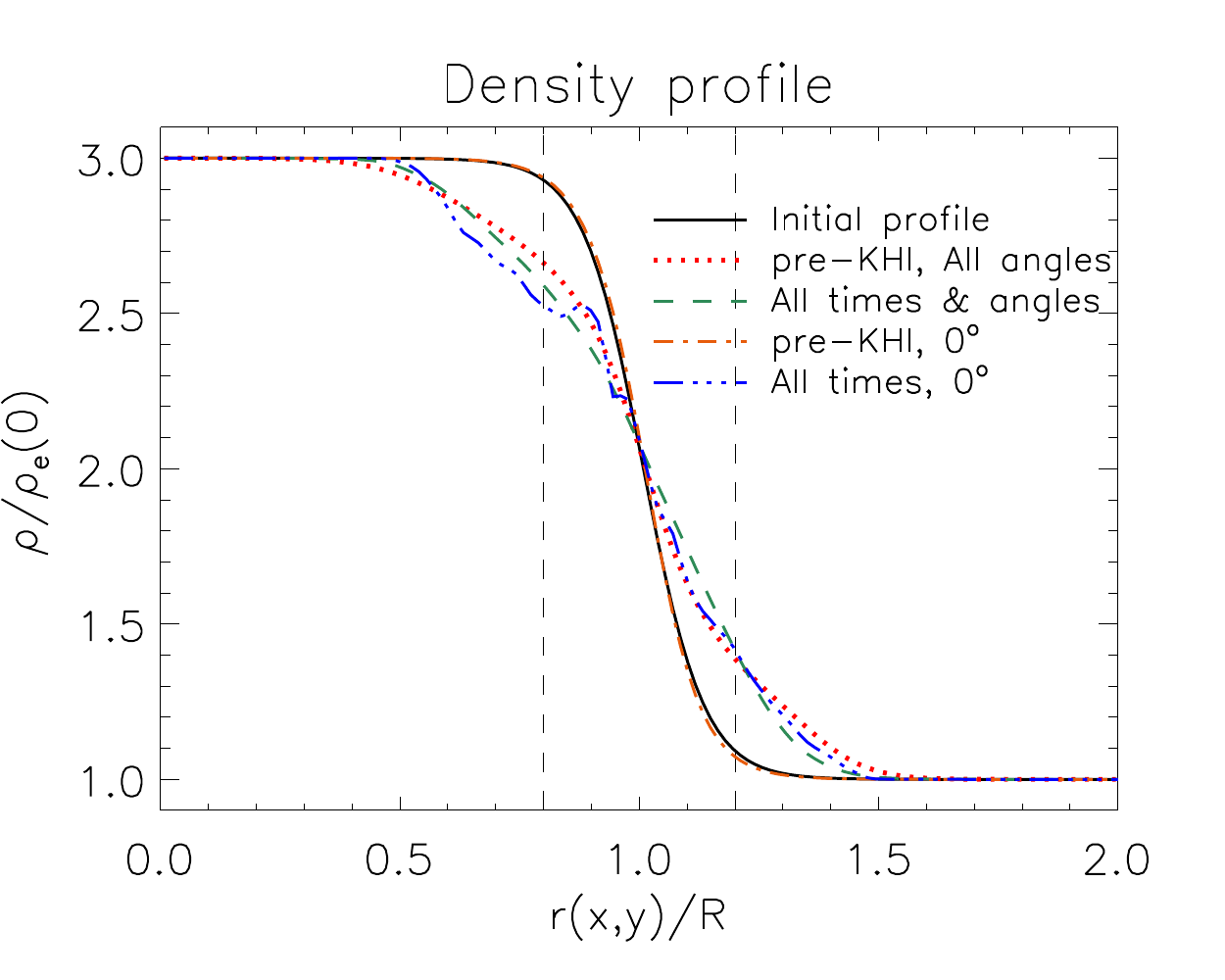}
\includegraphics[scale=0.35]{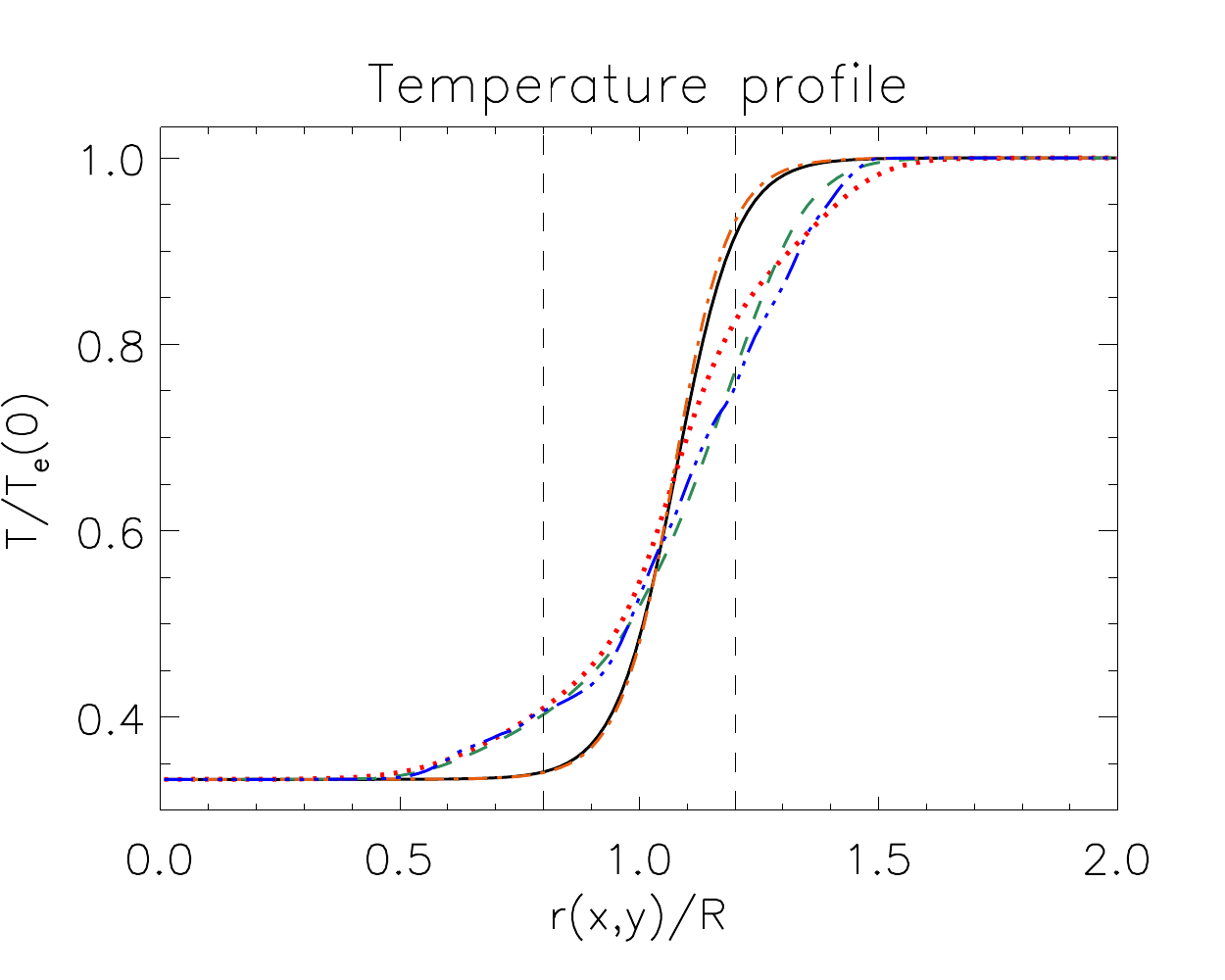}
\end{array}$
\caption{Initial density (\textit{left}) and temperature (\textit{right}) profiles (solid curves) across a perpendicular cross-section of the loop corresponding to \textit{model~1}. The boundary layer width $\ell$ is roughly $0.4~R$. Overlaid, their respective averages over all angles and times prior to KHI onset (dotted curves), the averages over all times and angles (dashed curves), the averages over times prior to KHI along the $y$-direction crossing the loop centre ($0^{\circ}$) in the Lagrangian frame of reference (dot-dashed curves), and the averages over the same direction but over all times (dot-dot-dashed curves).The vertical dashed lines denote, roughly, the start and end of the initial boundary layer. 
\label{fig2}}
\end{center}
\end{figure}

In \textit{model~1}, our 3D MHD numerical model is the same as in \citet{Antolin_2014ApJ...787L..22A}, where we take a loop with a density and temperature contrast with respect to the ambient low-$\beta$ coronal atmosphere. The loop is initially in hydrostatic equilibrium and has density and temperature ratios $\rho_i/\rho_e=3$ and $T_i/T_e=1/3$, respectively, where the index $i$ ($e$) denotes internal (external) values. The magnetic field is uniformly set to $B_0=22.8$~G. We take initially $T_i=10^6~$K, $\rho_i=3\times10^9 \mu m_{p}$~g~cm$^{-3}$ ($\mu=0.5$ and $m_p$ is the proton mass) and the density profile across the loop is set as follows and is shown in Fig.~\ref{fig2}:
\begin{equation}\label{eq1}
\rho(x,y) = \rho_e+(\rho_i-\rho_e)\zeta(x,y),
\end{equation}
where
\begin{equation}\label{eq2}
\zeta(x,y) = \frac{1}{2}(1-\tanh(b(r(x,y)-1))).
\end{equation}
In these equations, $x$ and $y$ denote the coordinates in the plane perpendicular to the loop axis, and $z$ is along its axis. The $r(x,y)=\sqrt{x^2+y^2}/R$ term denotes the normalised radius of the loop at position $(x,y)$ and $b$ sets the width of the boundary layer (the loop shell). In our model we set $b=16$, leading to $\ell/R\approx0.4$, where $\ell$ denotes the width of the boundary layer and $R$ is the radius of the loop (see Fig.~\ref{fig1}). The length $L$ of the loop is $200~R$, and we set $R=1~$Mm.

In the second numerical setup, \textit{model 2}, the temperature is uniformly set throughout to $T=7\times10^5$~K. The density contrast is the same as in the first model. The magnetic field varies with the inverse density profile in order to keep hydrostatic pressure balance (with a minimal variation from $B_e=11~$G to $B_i$=10.8~G). 

The loop in both models is subject to a perturbation mimicking a fundamental kink mode (longitudinal wavenumber $kR=\pi R/L\approx$0.015), by imposing initially a perturbation along the loop for the transverse $x-$velocity component, according to $v_{x}(x,y,z) = v_0 \cos(\pi z/L)\zeta(x,y)$, where $v_0$ is the initial amplitude. The corresponding kink phase speed is $c_k=\sqrt{(\rho_{i}v_{A_{i}}^2+\rho_{e}v_{A_{e}}^2)/(\rho_i+\rho_e)}\approx1574~$km~s$^{-1}$ and $752$~km~s$^{-1}$, respectively for \textit{models~1} and \textit{2}, where $v_{A_{i}}$ and $v_{A_{e}}$ denote the internal and external Alfv\'en speeds, equal to 1285~km~s$^{-1}$ and 2225~km~s$^{-1}$, respectively, in \textit{model~1} and 608~km~s$^{-1}$ and 1075~km~s$^{-1}$ in \textit{model~2}. Here we present results for $v_0 =0.05~c_s$, with $c_s$ the external sound speed in each model. This corresponds to $v_0=$15~km~s$^{-1}$ and 7~km~s$^{-1}$ in \textit{models~1} and \textit{2}, respectively. 

\subsection{Numerical setup}
We perform the 3D MHD simulation described above with the CIP-MOCCT scheme \citep{Kudoh_1999_CFD.8}. The MHD equations are solved, excluding gravity and loop curvature, which are second order factors for the present work. Furthermore, the effects of radiative cooling and thermal conduction are also neglected, as they are expected to be unimportant due to their longer timescales compared to that set by the kink wave \citep[see also][]{Cargill_2016ApJ...823...31C}. 

The numerical box is 512 $\times$ 256 $\times$ 50 points in the $x, y$ and $z$ directions respectively. Thanks to the symmetric properties of the kink mode only half the plane in $y$ and half the loop are modelled (from $z=0$ to $z=100~$R), and we set symmetric boundary conditions in all boundary planes except for $x$, where periodic boundary conditions are imposed. In order to minimise the influence from side boundary conditions (along $x$ and $y$), the spatial grids in $x$ and $y$ are non-uniform, with exponentially increasing values for distances beyond the maximum displacement. The maximum distance in $x$ and $y$ from the centre is $\approx 16~R$. The spatial resolution at the loop's location is 0.0156~R$=15.6~$km. The code includes explicit resistive and viscous coefficients, which are however set to very low values in the current simulation. The simulation can therefore be considered ideal to a certain extent. However, from a parameter study, we estimate that the effective Reynolds and Lundquist numbers in the code are of the order of $10^4-10^5$. We have checked that the energy is conserved to high accuracy in the whole numerical box in the current simulation.

\section{Results - general characteristics}\label{results}

\subsection{Two mechanisms combined: resonant absorption and the Kelvin-Helmholtz instability}

Following the initial perturbation with amplitude $v_0=15~$km~s$^{-1}$ and $v_0=7~$km~s$^{-1}$ for \textit{models~1} and \textit{2} respectively, the loop starts oscillating with a period $P=255~$s and $P=530~$s (respectively), closely corresponding to the period of the fundamental mode $2L/c{_k}$. The maximum displacement of the loop is 440~km and 430~km, that is, $0.44~R$ and $0.43~R$, in each model, respectively. As expected, resonant absorption rapidly sets in in the loop boundary. Energy is then transferred from global transverse oscillations to local azimuthal oscillations within the loop boundary, clearly visible in the $z-$component of the vorticity in Movie~1.

\begin{figure}[!ht]
\begin{center}
$\begin{array}{cc}
\includegraphics[scale=0.33]{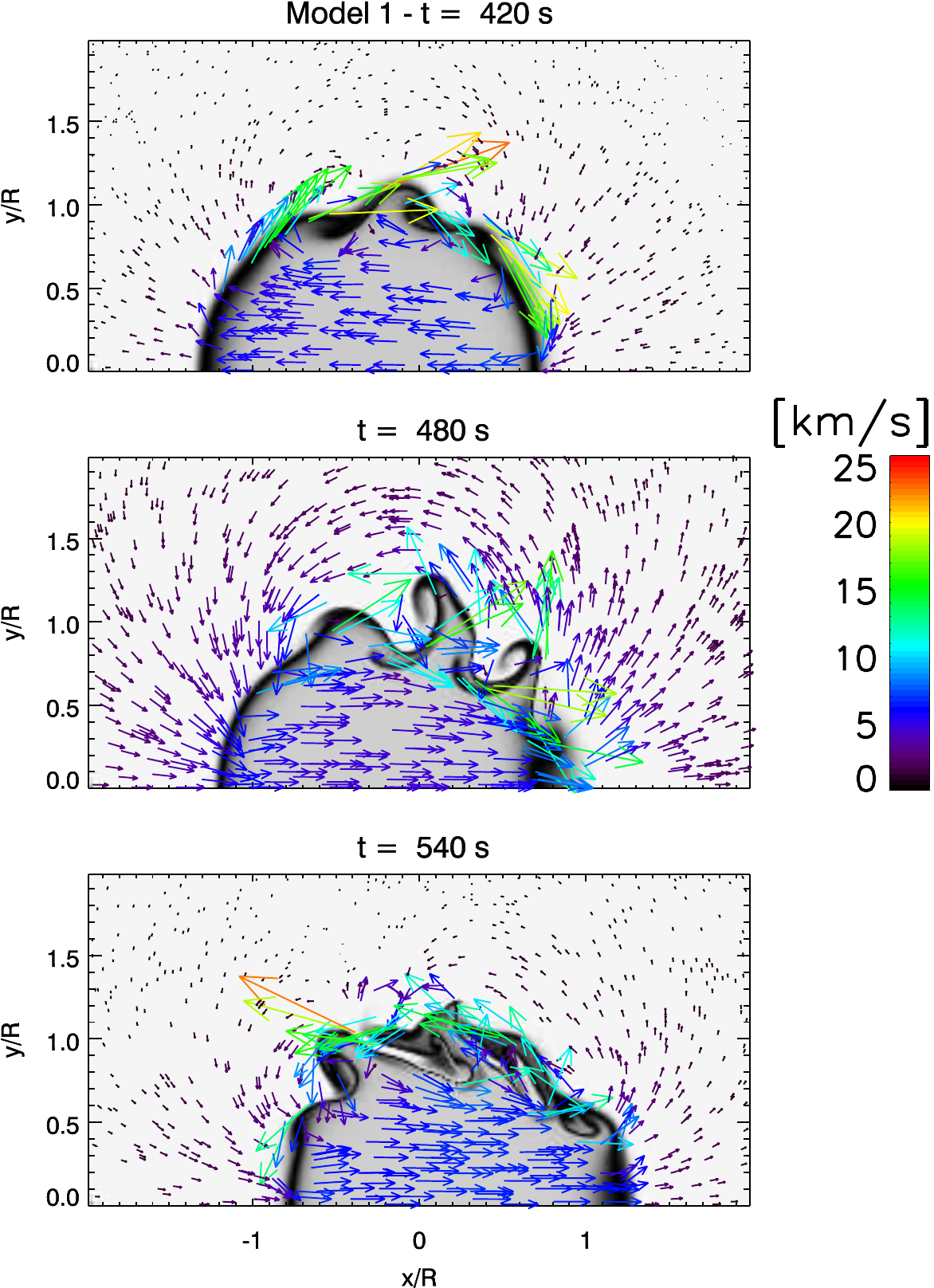} &
\includegraphics[scale=0.33]{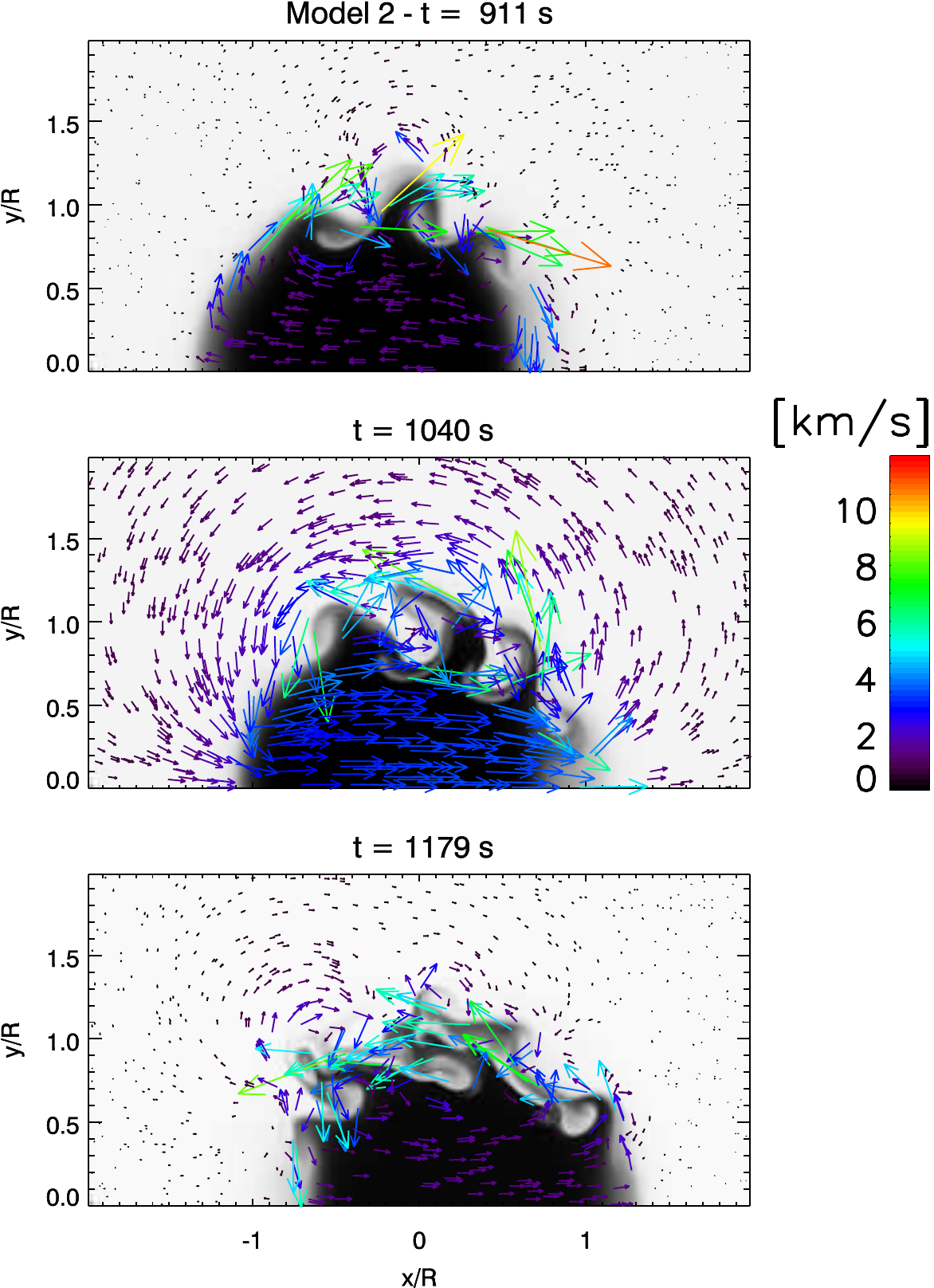}
\end{array}$
\caption{Snapshots of the emission line flux (in inverted grey-scale colours) in \ion{Fe}{12}~193  for \textit{model~1} (left panels) and in \ion{Fe}{9}~171 for \textit{model~2} (right panels), with the velocity field overlaid in rainbow colours. The snapshots follow the formation of the first KHI vortices (named TWIKH rolls). The colour scale for the velocity field incorporates the extrema over the entire simulation. 
\label{fig3}}
\end{center}
\end{figure}

In addition to resonant absorption a second mechanism sets in. As mentioned in the introduction, the kink mode produces a velocity shear at the boundary of the flux tube, between the dipole-like external flow and the purely transverse motion of the kink mode. Both motions are part of the kink mode, the former (dipole-like flow) being produced by the quadrupolar terms in the wave equation \citep{Goossens_2014ApJ...788....9G,Yuan_2016ApJS..223...23Y}. In the presence of a density gradient, the shear is amplified by the presence of the (azimuthal) resonant flow. This shear generates the Kelvin-Helmholtz instability (KHI), as recently demonstrated analytically by \citet{Zaqarashvili_2015ApJ...813..123Z}. In Fig.~\ref{fig3}, which shows snapshots of the emissivity of a cross-section of half the loop at the apex, we can see the formation of the initial vortices for both models (see also Movie~1). Three vortices appear first, and half a period later 4 vortices, indicating that $m=3$ and 4 are the most unstable modes, which matches theoretical results \citep[see Eq.~58 in][]{Zaqarashvili_2015ApJ...813..123Z}. 

\begin{figure}[!ht]
\begin{center}
\includegraphics[scale=0.4]{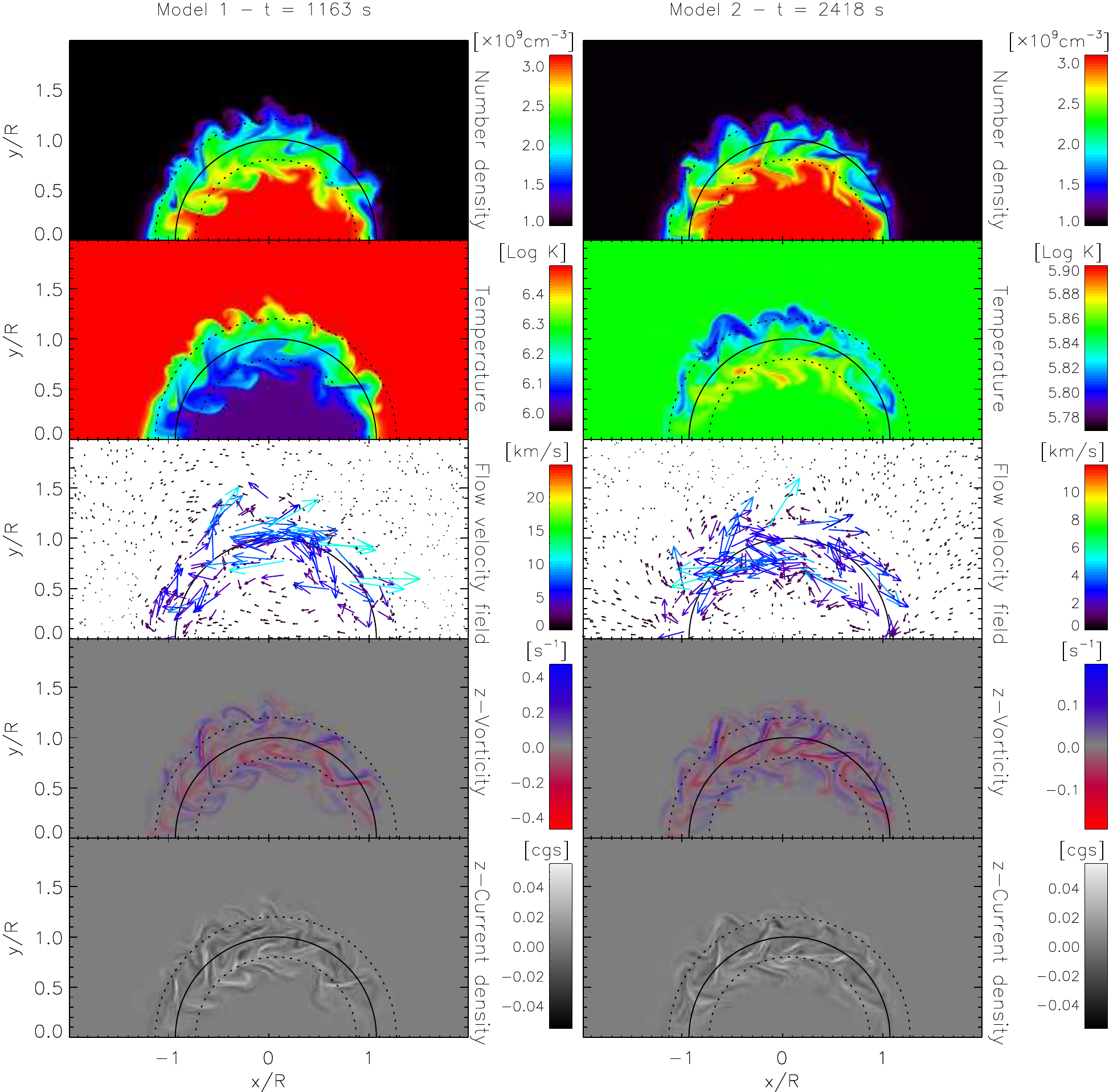}
\caption{A snapshot of the simulations in \textit{model~1} (left panels) and \textit{model~2} (right panels) after roughly 4 periods. From top to bottom we have the number density, temperature (logarithm), flow velocity field, and the longitudinal ($z-$) components of the vorticity and current density. We overlay in black the initial radius of the loop (solid curve) and the minimum and maximum extent of the boundary layer (dashed curves). An animation of this figure (Movie~1) is available online.
\label{fig4}}
\end{center}
\end{figure}

The KHI broadens the initial boundary layer, the extent of which can be seen in Fig.~\ref{fig2}. In this figure we plot the density and temperature profiles averaged over time and angles around the cylindrical cross-section at the centre of the loop, for times prior to the KHI onset and also over the entire simulation. Prior to the onset of the KHI the flux tube suffers a periodic compression and rarefaction at the loop boundary, leading to a deformation of the loop's cross-section. The compression and rarefaction occur, respectively, at the head and tail of the flux tube (along the axis of oscillation). This squashing effect is produced by the combination of fluting modes and the inertia produced by the wave (which acts differently on the loop's core and shell), and is completely reversible (the boundary shape in the $x-$direction goes back to the original shape each time the flux tube passes through the initial position). On the other hand, the KHI has an irreversible smoothing effect over all angles and is maximal at the sides of the flux tube that are perpendicular to the direction of oscillation. This is expected since these are the locations of maximum velocity shear. At the end of the simulation, the centre of the loop is roughly back to the initial position and the loop core is squashed in the oscillation direction. The KHI therefore leads to elliptically shaped loop cores with boundary layers that have roughly linear density gradients and with major axis along the direction of oscillation, as seen in Fig.~\ref{fig4}.

\subsection{Compressive vortices and the thread-like structure in EUV emission lines}\label{compvortex}

The KHI produces vortices (TWIKH rolls) in the transverse cross-section along most of the flux tube's length. Large scale vortices are generated first (at the location of the resonance), which rapidly break-up into smaller and smaller vortices. As shown in Fig.~\ref{fig4} such vortices mix the external ambient plasma with the internal loop plasma, and produce current sheets and locations of viscous and resistive dissipation. Comparing the results from \textit{model~1} and from \textit{model~2} suggest that most of the change in loop temperature produced in the present scenario seems to come from the mixing with the external plasma and not from actual wave dissipation. As shown by Fig.~\ref{fig4} (and Movie~1), the region corresponding to positive temperature change in \textit{model~2} is the inner shell region, where the density is reduced due to the KHI, while the outer shell region, which gets denser, gets colder. The changes of temperature are thus mainly due to a change in density and are thus mostly adiabatic rather than linked to wave dissipation \citep{Karampelas_2017}.

The vortices (or roll-ups in 3D) have a fast MHD mode nature in the sense that they are able to compress the plasma and therefore significantly deform the density and temperature structure in the transverse plane of the flux tube, similar to vortices generated through Alfv\'enic vortex shedding \citep{Gruszecki_2010PhRvL.105e5004G}. This is especially true in chromospheric conditions \citep{Antolin_2015ApJ...809...72A} or for mild amplitude perturbations above $0.01v_{A_i}$. Such vortices are therefore regions of enhanced emissivity. This can be clearly seen in Fig.~\ref{fig3}, where the TWIKH rolls have higher intensity (dark in the colour inverted image). Due to the optically thin nature of the corona, the emission from these vortices adds up along the LOS, resulting in clear strand-like structure in EUV intensity images of the loop, as first shown in \citet{Antolin_2014ApJ...787L..22A,Antolin_2015ApJ...809...72A}.

\section{Forward modelling the uniform and non-uniform temperature models}\label{forwardresults}

\subsection{Emission lines, LOS and instrumental setup}\label{forward}

To obtain observable quantities, the results from the numerical simulations are forward modelled using  the FoMo code \citep{VanDoorsselaere_0.3389/fspas.2016.00004}, which calculates the optically thin emission of coronal lines based on the CHIANTI atomic database \citep{Dere_etal_2009AA...498..915D}.

We choose the \ion{Fe}{9}~171 and \ion{Fe}{12}~193 emission lines, which have rest wavelengths at 171.073~\AA~and 193.509~\AA, and maximum formation temperatures of $\log T=5.93$ and $\log T=6.19$, respectively. For \textit{model~1}, the \ion{Fe}{9}~171 line is more tuned to detect the plasma response at the core of the loop, while the \ion{Fe}{12}~193 line is more sensitive to the hotter plasma near the boundaries. For \textit{model~2} the temperature in and around the loop does not deviate much from the initial temperature throughout the simulation, which is close to the formation temperature of the \ion{Fe}{9} line. For that model we therefore only use the \ion{Fe}{9} line for the forward modelling. From here on, we will refer to the \ion{Fe}{9}~171 line as the core line, and the \ion{Fe}{12}~193 line as the boundary line. 

We define the LOS angle such that $0^{\circ}$ is parallel to the positive y-axis, and $90^{\circ}$ is parallel to the positive $x$-axis, which is the axis of oscillation. This is shown schematically in Fig.~\ref{fig1}. For correct comparison with observations with a given instrument with resolving power of X (defined as the FWHM of the Point-Spread Function - PSF - and we take the plate-scale equal to half the spatial resolution, unless explicitly stated) we degrade the original spatial resolution of the numerical model by first convolving the image of interest with a Gaussian with FWHM of X. We then resample the data according to the specific pixel size of the target instrument and add photon noise (Poisson distributed).  

In the forward modelling we have also considered various spectral and temporal resolutions. For a given spectral resolution, we convolve the numerical results with a PSF in the spectral dimension of the same size. We then resample the numerical results in the spectral and temporal dimensions with the target spectral resolution and cadence. Besides adding photon noise we also add a $5\%$ random fluctuation at each wavelength position over the profile. 

Unlike \textit{model~1}, for which the ambient coronal environment has little emissivity in the selected emission lines, \textit{model~2} suffers from background emission from plasma at rest, due to the similar temperature inside and outside the flux tube. We therefore opt to subtract this background contribution prior to analysis by eliminating first the effect of having a square numerical box, which makes different LOS rays have different lengths. For this, we add artificial pixels with the same emission as a pixel with a background profile at rest, so that any LOS ray has equal length. A background subtraction is subsequently performed on every pixel, which now acts equally along any LOS. 

\subsection{Imaging signatures}

\begin{figure}[!ht]
\begin{center}
$\begin{array}{c}
\includegraphics[scale=0.44]{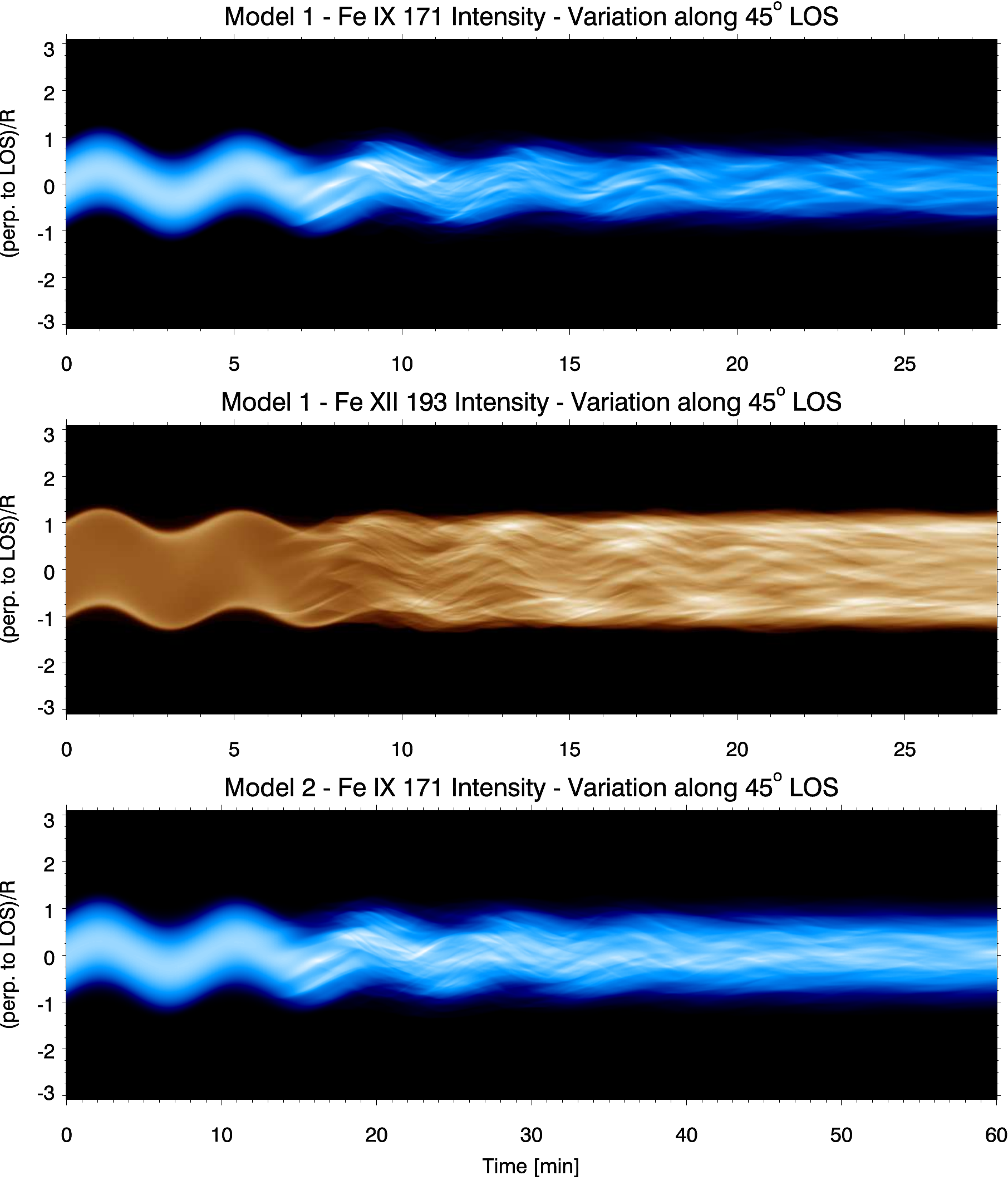}
\end{array}$
\caption{Time-distance diagrams of the intensities in the \ion{Fe}{9} (\textit{top panel}) and \ion{Fe}{12} (\textit{middle panel}) lines of \textit{model~1} and \ion{Fe}{9} line (\textit{bottom panel}) of \textit{model~2} for a slit placed perpendicularly to the loop at the apex and with a LOS angle of $45^{\circ}$ and at numerical (highest) spatial resolution. 
\label{fig5}}
\end{center}
\end{figure}

\begin{figure}[!ht]
\begin{center}
$\begin{array}{c}
\includegraphics[scale=0.6]{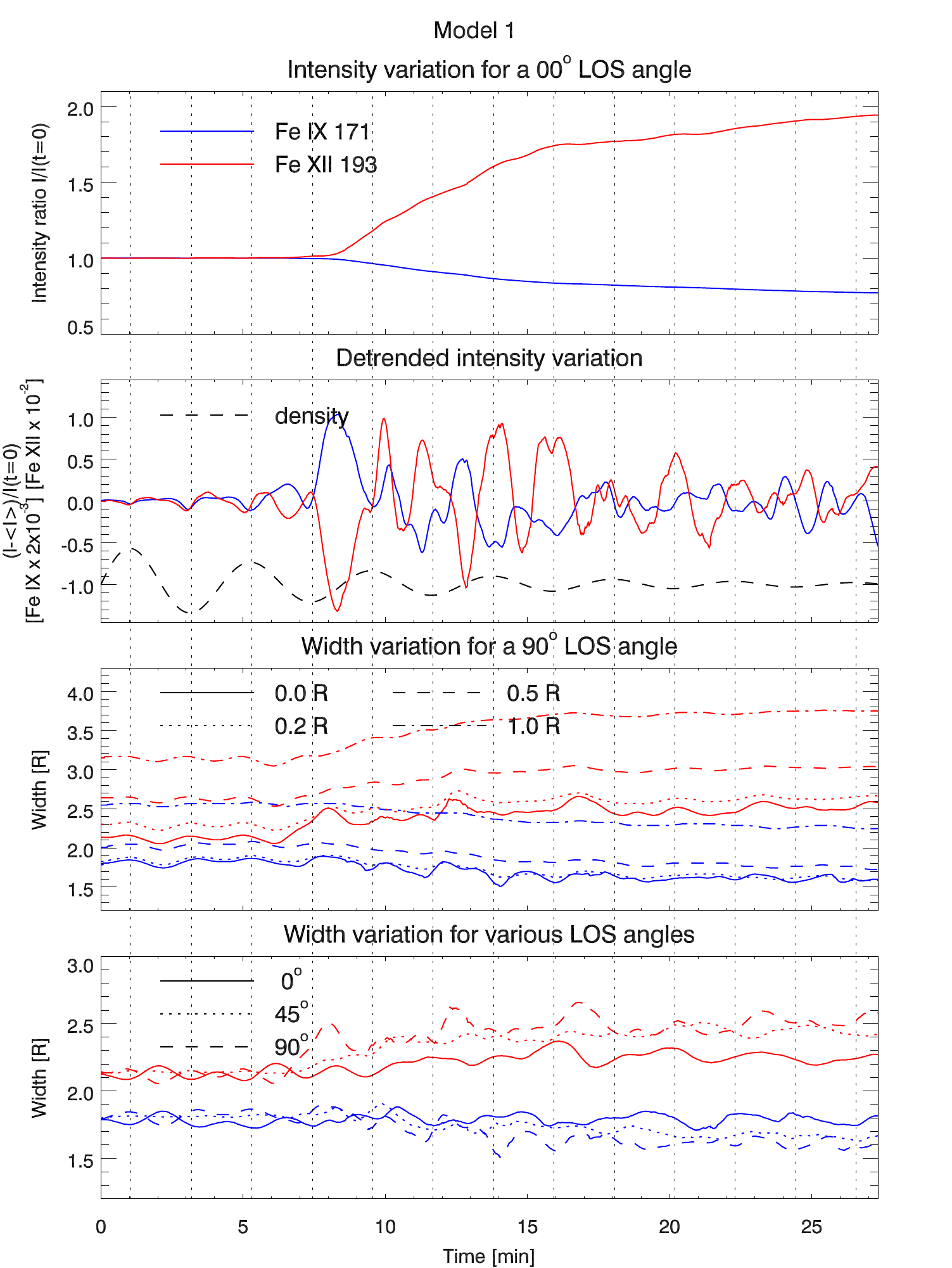}
\end{array}$
\caption{For all panels we show in red and blue colours the temporal variation of quantities  associated with the \ion{Fe}{9} and \ion{Fe}{12} lines respectively. \textit{Top panel:} Variation of the intensity integrated over a slit placed perpendicular to the loop at the apex and at a LOS angle of $0^{\circ}$. \textit{Second panel from top:} De-trended and normalised intensity shown in the top panel (note that the \ion{Fe}{9} intensity has been multiplied by 5 to fit in the same scale). The fit to the density is overlaid with a dashed curve with an ad-hoc scale. The fit is done by first degrading the spatial resolution to $1~R$ for the density cross-section at the (previously defined) slit location, then fitting a Gaussian profile at each time (from which the centroid is calculated), and fitting an exponentially damped cosine to the centroid locations. \textit{Third panel from top:} Variation of loop width calculated from each intensity line for a slit placed perpendicular to the loop at the apex, at an angle of $90^{\circ}$ and for different spatial resolutions. The loop boundaries are defined by the locations where the loop intensity increases above 20~\% of the initial maximum intensity in each line. \textit{Bottom panel:} Variation of the loop width for different LOS angles and for a spatial resolution of $0~R$ (highest). The vertical dotted lines correspond to the extrema of the density fit oscillation.
\label{fig6}}
\end{center}
\end{figure}

\begin{figure}[!ht]
\begin{center}
$\begin{array}{c}
\includegraphics[scale=0.6]{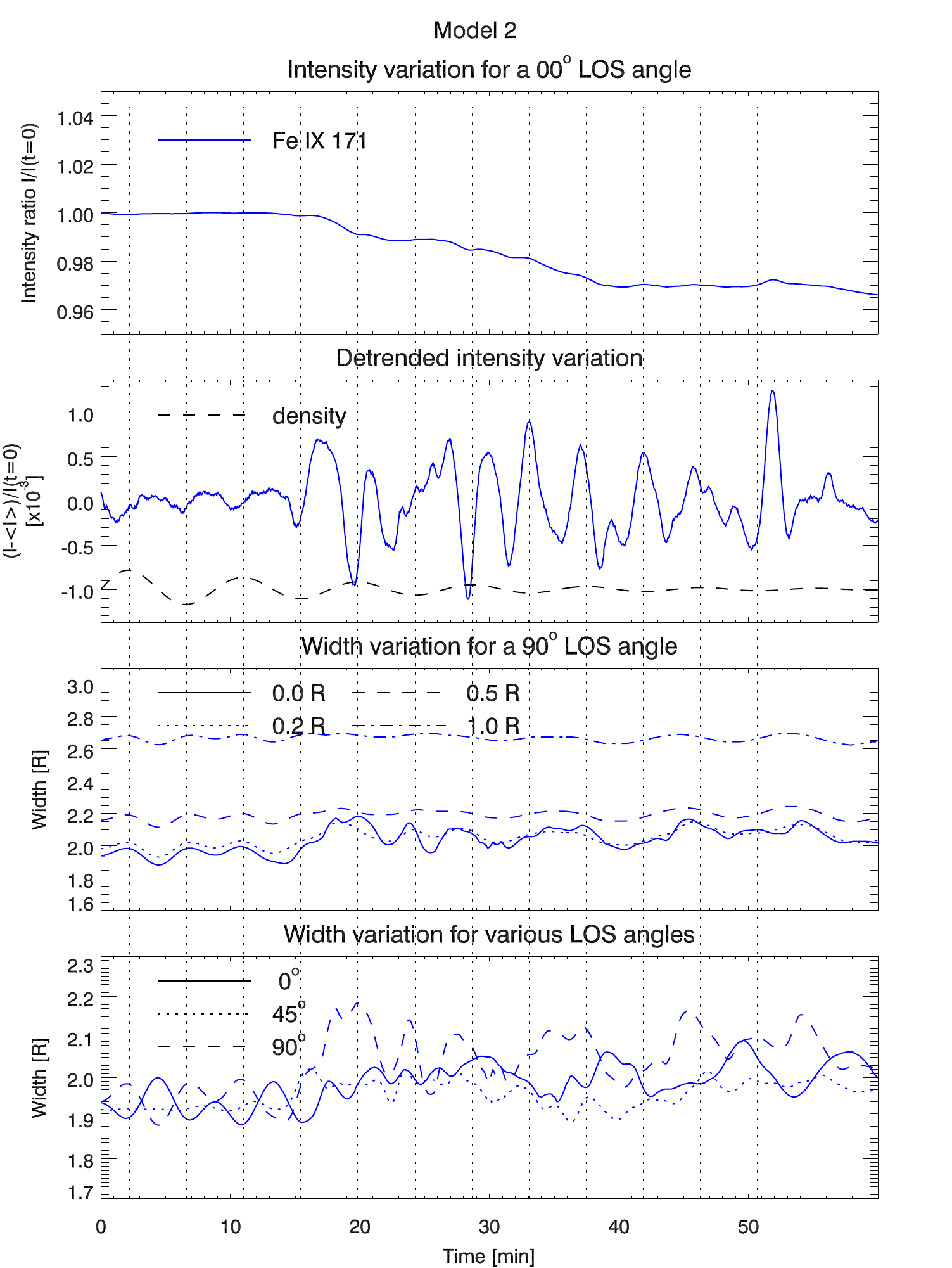}
\end{array}$
\caption{Similar to Fig.~\ref{fig6} but for the \ion{Fe}{9} de-trended intensity of \textit{model~2}.
\label{fig7}}
\end{center}
\end{figure}

Figure~\ref{fig5} shows the time-distance diagram of the intensity in \ion{Fe}{9}~171 (for \textit{models~1} and \textit{2}) and \ion{Fe}{12}~193 (only for \textit{model~1}) for an artificial slit placed transverse to the loop at its apex and for a LOS angle of 45$^{\circ}$ (see section~\ref{forward} for the convention on the LOS angle). Since the \ion{Fe}{9}~171 line is more tuned to detect the plasma response at the core of the loop in \textit{model~1}, while the \ion{Fe}{12}~193 line is more sensitive to the hotter plasma near the boundaries, we can see that the loop diameter is slightly larger in the boundary line at time $t=0$ as well as throughout the evolution. Correspondingly, as we will see clearly later on, the width of the loop in \textit{model~2} is wider than that in \textit{model~1}.

Bright oscillatory strand-like structures are noticeable in both emission lines and in both models after about 2 periods, corresponding to the formation of TWIKH rolls. The amount of strand-like structure is largest in the boundary line of \textit{model~1}. In the core line we can see that the loop in \textit{model~2} shows a significantly larger amount of this fine structure than the loop in \textit{model~1}, as the uniform temperature of \textit{model~2} captures the TWIKH rolls in the cooler 171 line. 

Although the overall damping is clear in \ion{Fe}{9}~171 for both models, the inner loop structure in the \ion{Fe}{12}~193 emission appears `decay-less', as explained in \citet{Antolin_2016ApJ...830L..22A}.

Another clear difference in the time-distance diagrams of Fig.~\ref{fig5} between the core and boundary lines is the presence of periodic intensity brightening only in the boundary line near the loop boundary after KHI onset. This periodic brightening is further seen in Fig.~\ref{fig5} to extend over time, lasting under a minute first and being almost continuous by the end of the simulation. In the first 2 periods prior to KHI onset a small periodic brightening can also be seen in the boundary line in the trailing edge of the loop close to times of maximum displacement. 

While the loop is observed to become thinner and dimmer in time for \textit{model~1} in \ion{Fe}{9}~171, in \ion{Fe}{12}~193 the loop becomes broader and brighter. The brightness increase and broadening occurs rapidly, in about one period after the onset of the KHI. In contrast, the intensity in \ion{Fe}{9}~171 in \textit{model~2} seems roughly constant on average. In Fig.~\ref{fig6} for \textit{model~1} and in Fig.~\ref{fig7} for \textit{model~2} we plot the evolution of the integrated intensity over the slit. We can see in \textit{model~1} (top panel of Fig.~\ref{fig6}) a clear intensity enhancement in the boundary line, occurring right after the onset of the KHI with a steep linear increase over 2 periods up to a factor of 1.8, followed by a gradual increase up to a factor of 2 in the remainder of the simulation. On the other hand, the core line in the same model shows a dimming to a weaker degree, but also in a 2-step behaviour, leading to an overall decrease of 20\%. The trend observed in the same line for \textit{model~2} is very similar to that of \textit{model~1} but with a much smaller magnitude of only $3\%-4\%$. This is because of the competing effect from adiabatic heating and cooling within the shell, as seen in Fig.~\ref{fig4}.

Similarly, on a relatively short timescale of about 2 periods after the KHI onset, the loop width increases, as shown in Fig.~\ref{fig2}. This evolution can be seen in Fig.~\ref{fig6} for \textit{model~1}, and in Fig.~\ref{fig7} for \textit{model~2}, for even low spatial resolution. After KHI onset, the loop width increases for the boundary line up to $17-24$~\%, while it decreases for the core line by a factor of $11-15$~\%. The variation in loop width for \textit{model~2} in the core line looks similar to that of the boundary line in \textit{model~1}. Indeed, although with a lower amplitude of only $3\%$, a general increase in loop width can be seen in the third panel of Fig.~\ref{fig7}, but only for high spatial resolution (below $0.5R$).

The small scale periodic brightening near the loop boundary has an impact on the overall intensity. Indeed, by summing the intensity across the flux tube and subtracting the (smoothed out) general trend of the intensity evolution we can discern multiple small oscillations in the intensity with a maximum of about 1.5\% and 0.2\% for \ion{Fe}{12} and \ion{Fe}{9}, respectively for \textit{model~1}, and $0.1\%$ in \ion{Fe}{9} for \textit{model~2}. This is shown in the second panel from the top in Figs.~\ref{fig6} and \ref{fig7}. Initially the intensity oscillations in both lines in \textit{model~1} are in-phase and they go out-of-phase as soon as the KHI sets in, suggestive of different mechanisms. The main intensity peaks in the boundary line are reached in general at times of maximum displacement (minimum velocity shear), and at times of minimum displacement (maximum velocity shear) for the core line. The time locations of these peaks are independent of the LOS.

Periodic oscillations in the width of the loop can also be seen for any line and model. At high resolution, the loop width variation over half a period can be as high as 25~\%, but decreases to 3~\% for a low spatial resolution of $1~R$. This effect has also been described in \citet{Yuan_2016ApJS..223...23Y}, although the origin of this behaviour is different in both studies. The bottom 2 panels in Fig.~\ref{fig6} for \textit{model~1} and in Fig.~\ref{fig7} for \textit{model~2} show the evolution of the loop width for different spatial resolutions and for different LOS angles. Oscillations in the loop width can be seen with double the periodicity of the kink mode, in-phase between both lines prior and after the KHI, although with some scatter. 

Prior to KHI onset these intensity (and loop width) oscillations are about a factor of 5 smaller, as expected from the highly incompressible nature of the kink mode \citep{VanDoorsselaere_2008ApJ...676L..73V, Goossens_2012ApJ...753..111G}. The small oscillations in intensity obtained during this time window are partly due to the quadrupolar terms in the wave solution characteristic of the kink mode, as described by \citet{Yuan_2016ApJS..223...23Y}, and to a minor degree to resonant absorption (which redistributes material in the boundary layer), and mainly due to the deformation of the flux tube from the combined effect of the inertia and fluting modes \citep{Ruderman_2010PhPl...17h2108R}, both of which produce an ellipse with maximum axis along the $90^{\circ}$ LOS at times of maximum displacement. All these effects affect the column depth (and are therefore dependent of the LOS angle) in a similar way for both lines, explaining the in-phase behaviour. 

The main features in Figs.~\ref{fig6} and \ref{fig7} are due to the TWIKH rolls, as discussed in section~\ref{compvortex}. For \textit{model~1} the dominant effect of the TWIKH rolls is the mixing of the internal and external plasma. This produces an increase of emission in the boundary line and a corresponding decrease in the core line, since the temperature of the plasma in the vortices shifts away from the maximum formation temperature of the core line and closer to that of the boundary line, thereby explaining the anti-phase behaviour in the de-trended intensity oscillation and the overall increase and decrease in the boundary and core lines, respectively. For \textit{model~2}, the intensity amplitude oscillations are mostly due to the compressive effect of the TWIKH rolls and are smaller than in \textit{model~1}. TWIKH rolls in this model generally achieve their greatest size at times of zero displacement. Furthermore, a vortex forming at the trailing side of the loop will be carried on top of the moving loop on the way back. Therefore, the vortex will be further away from the plane of oscillation at times of zero displacement, thus increasing the loop width (particularly for a LOS of 90$^{\circ}$), and leading mostly to an in-phase modulation in both spectral lines. This effect also makes the loop width variation dependent on the LOS angle, as can be seen in Figs.~\ref{fig6} and \ref{fig7} (check the out-of-phase behaviour between the $0^{\circ}$ LOS case and the $90^{\circ}$ LOS case for both lines). This behaviour is not obtained when considering only the linear (quadrupolar) terms of the wave equation \citep{Yuan_2016ApJS..223...23Y}.

\begin{figure}[!ht]
\begin{center}
$\begin{array}{c}
\vspace{-1.cm}\includegraphics[scale=0.57]{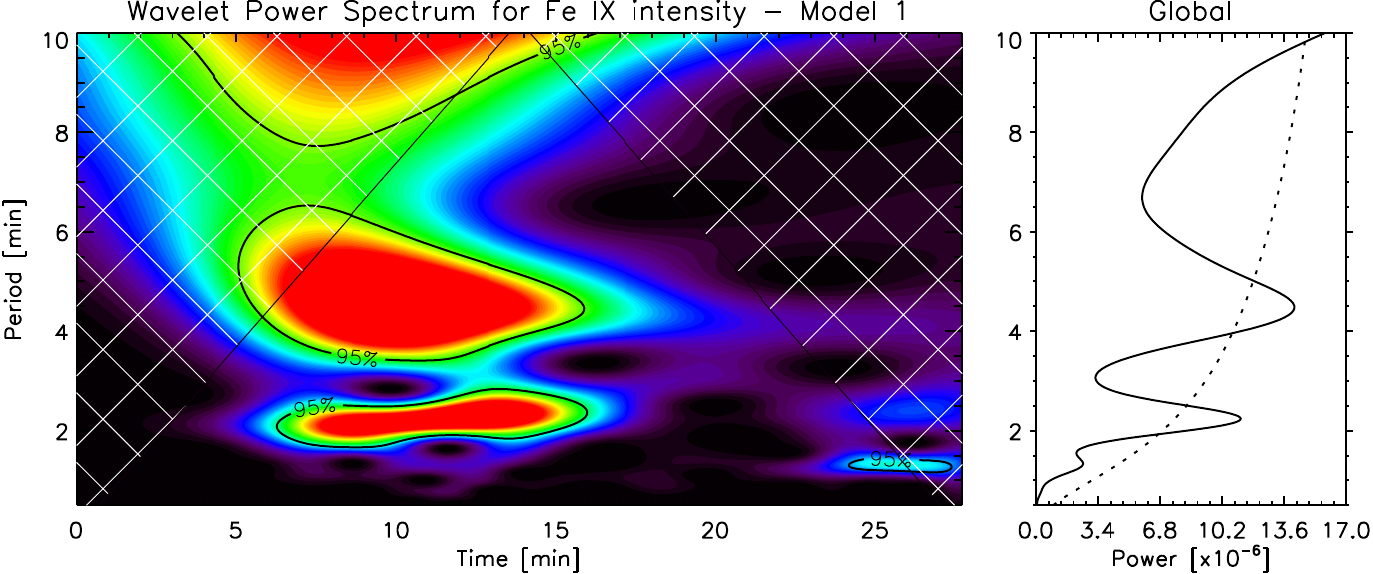} \\
\vspace{-1.cm}
\includegraphics[scale=0.57]{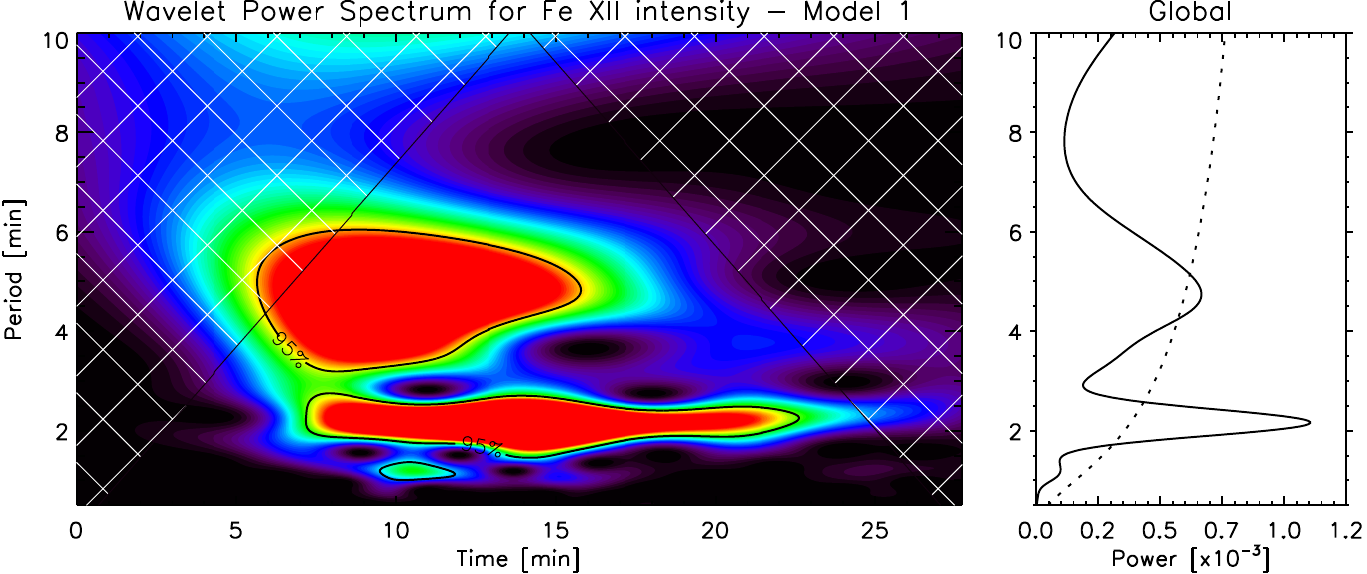} \\
\includegraphics[scale=0.57]{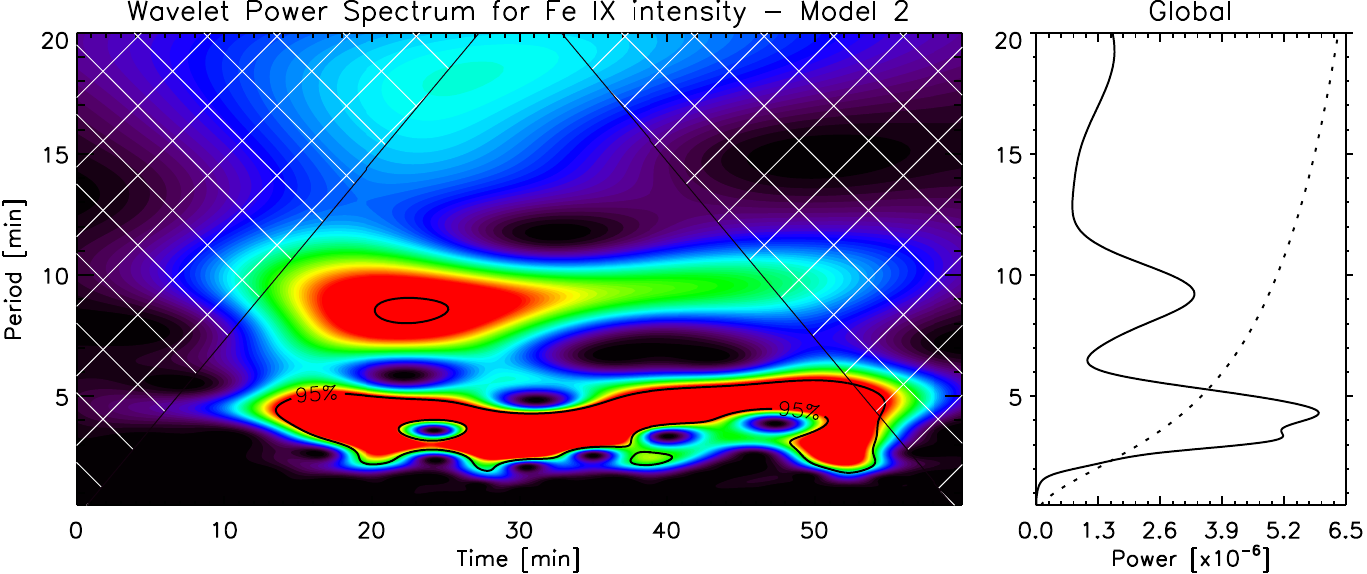}
\end{array}$
\caption{Wavelet analysis for \ion{Fe}{9} (top panel) and \ion{Fe}{12} (middle panel) de-trended intensity lines in \textit{model~1}, and \ion{Fe}{9} (bottom panel) in \textit{model~2} \citep[computed with the][software]{Torrence_1998BAMS...79...61T}. The significance level shown with thick contours encloses regions of greater than 95\% confidence, assuming a red-noise process as background spectrum (we estimate the lag-1 coefficient of 0.9). Cross-hatched regions indicate the cone of influence, where edge effects become important. 
\label{fig8}}
\end{center}
\end{figure}

Figures~\ref{fig8} shows the wavelet analysis for the de-trended intensity time series for both lines in \textit{model~1} and the core line in \textit{model~2}, in which we can see the prevalence of mainly two periodicities, one at the global kink period (roughly 4.5~min and 9~min in \textit{models~1} and \textit{2}, respectively) and a similar or stronger peak at half that periodicity, appearing when the KHI sets in. The power is also more extended in time for the boundary line in \textit{model~1} and the core line in \textit{model~2}. Interestingly, the power of the peak corresponding to the global kink mode is wider in frequency in \textit{model~1} for both lines, having roughly twice the width of the strongest peak at half the period. On the other hand, the width of both peaks in \textit{model~2} is comparable. 

\subsection{Spectral signatures - phase drifts and ragged transitions}\label{spectral}

\begin{figure}[!ht]
\begin{center}
$\begin{array}{c}
\includegraphics[scale=0.43]{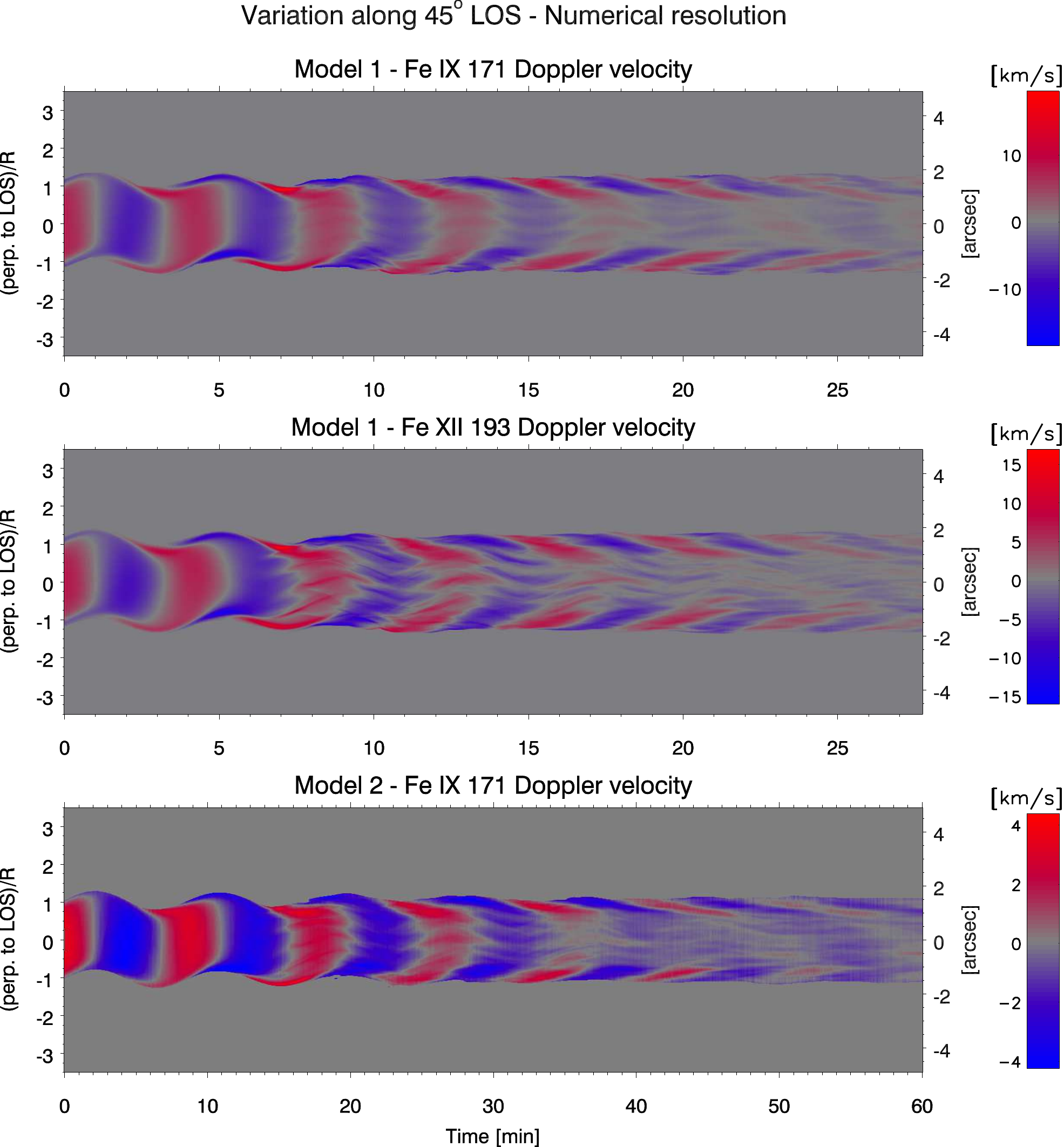}
\end{array}$
\caption{Doppler velocity in the \ion{Fe}{9} (\textit{top panel}) and \ion{Fe}{12} (\textit{middle panel}) lines of \textit{model~1} and \ion{Fe}{9} line (\textit{bottom panel}) for \textit{model~2} for a slit placed perpendicularly to the loop at the apex and with a LOS angle of $45^{\circ}$. The Doppler velocity is calculated by fitting a single Gaussian over the intensity profile for each pixel with an integrated intensity (over wavelength) above 15\% of the maximum integrated intensity at $t=0$.
\label{fig9}}
\end{center}
\end{figure}

Many of the observed features in imaging data can be further understood from the spectral diagnostics. Figure~\ref{fig9} shows the Doppler velocity in both emission lines in \textit{model~1} and in the core line for \textit{model~2} for the numerical resolution and a LOS of $45^{\circ}$. The initial blue- and redshifts at the centre of the flux tube correspond to the global kink mode and this Doppler oscillation is roughly 90$^{\circ}$ out-of-phase with the transverse (plane-of-the-sky, POS) motion, as expected \citep{Goossens_2014ApJ...788....9G}. At the loop boundary, corresponding to the top and bottom (thin) boundary regions in the figure, alternated thin `envelopes' of opposite Doppler shift with respect to the core are observed, immediately from the start of the simulation. These initial thin envelopes correspond to the response of the external plasma to the kink mode, which moves azimuthally around the flux tube, in opposite direction (dipole return flows). It is worth specifying that the calculation of the Doppler component over the spectral profile has been done at locations with an intensity above 15~\% that of the maximum initial intensity. For \textit{model~2}, where the external and internal plasma have the same temperature, we have opted to subtract the background contribution from plasma at rest, as described in section~\ref{forward}. This procedure also affects the intensity of the features at the boundary of the flux tube, thereby reducing the contribution from the quadrupolar terms. Correspondingly, the thin envelopes in Fig.~\ref{fig9} are much reduced for \textit{model~2}. Also, this procedure reduces the maximum detected Doppler velocity for that model, which should be similar or larger than the initial kink amplitude.

Over the first 2 periods an increase of the Doppler velocity amplitude in the thin envelopes (the loop boundary) is clearly seen in both lines and both models. This effect corresponds to resonant absorption. Accordingly, the largest Doppler components are the azimuthal resonant flows at the boundary of the flux tube, which are, however, confined to the very thin boundary layers until that moment in time (and confined therefore to regions of low emissivity, especially in the core line). When the KHI sets in, vortices are generated around the boundary, which carry similar characteristics as the resonant flow, especially its magnitude. 

The direction of motion of the vortices will also be partly defined by the direction of the resonant flow. Initially, the TWIKH rolls move opposite to the loop core, as can be seen in Fig.~\ref{fig3}. This produces the observed jagged, irregular transitions between the blueshift and the redshift regions seen in both the core and boundary lines in Fig.~\ref{fig9}, where in many locations the blue and redshifts appear interwoven. As the vortices grow, their crests continue moving with the resonant flow opposite to the loop core, while their troughs now move in the same direction as the loop core. This motion leads partly to the observed criss-crossing of features in Fig.~\ref{fig5}, and leads also to the generation of the next set of vortices (see the middle panel of Fig.~\ref{fig3}). Lastly, when the crests break they have maximum emissivity (especially in the boundary line) and have now the largest (azimuthal) velocity amplitudes. This effect produces a net 90$^{\circ}$ to 180$^{\circ}$ phase difference between the POS motion and the LOS velocity. This effect has been initially reported in \citet{Antolin_2015ApJ...809...72A} for the prominence case. Such phase difference has also been observed with \textit{IRIS} during a flare, by \citet{Brannon_2015ApJ...810....4B}, providing an alternative interpretation in terms of TWIKH rolls. The significant amplitude of the vortices also produces a filamentary structure in the Doppler images of Fig.~\ref{fig9}. This is especially clear in the boundary line since in \textit{model~1} the vortices are on average hotter than the loop core. 

The vortices rapidly break-up into smaller vortices, while new vortices emerge and become dominant. The emergence of new vortices on top of old ones (moving in different directions), combined with the damping of the kink mode, produces a fading of the Doppler signal towards the loop centre. This fading is not observed towards the edges. Since the LOS is tangential to the resonant flow there, most vortices along that LOS move coherently, and with increasing amplitude from the resonance. 

As the loop boundary is broadened due to the vortices, the density transition, and hence the change in phase speed, between the core and the external medium becomes flatter. This can be seen as arrow shaped structures in the Doppler images, which become flatter as time passes due to phase mixing. 

\subsection{Line broadening - boundary enhancement}\label{linebroad}

\begin{figure}[!ht]
\begin{center}
$\begin{array}{c}
\includegraphics[scale=0.43]{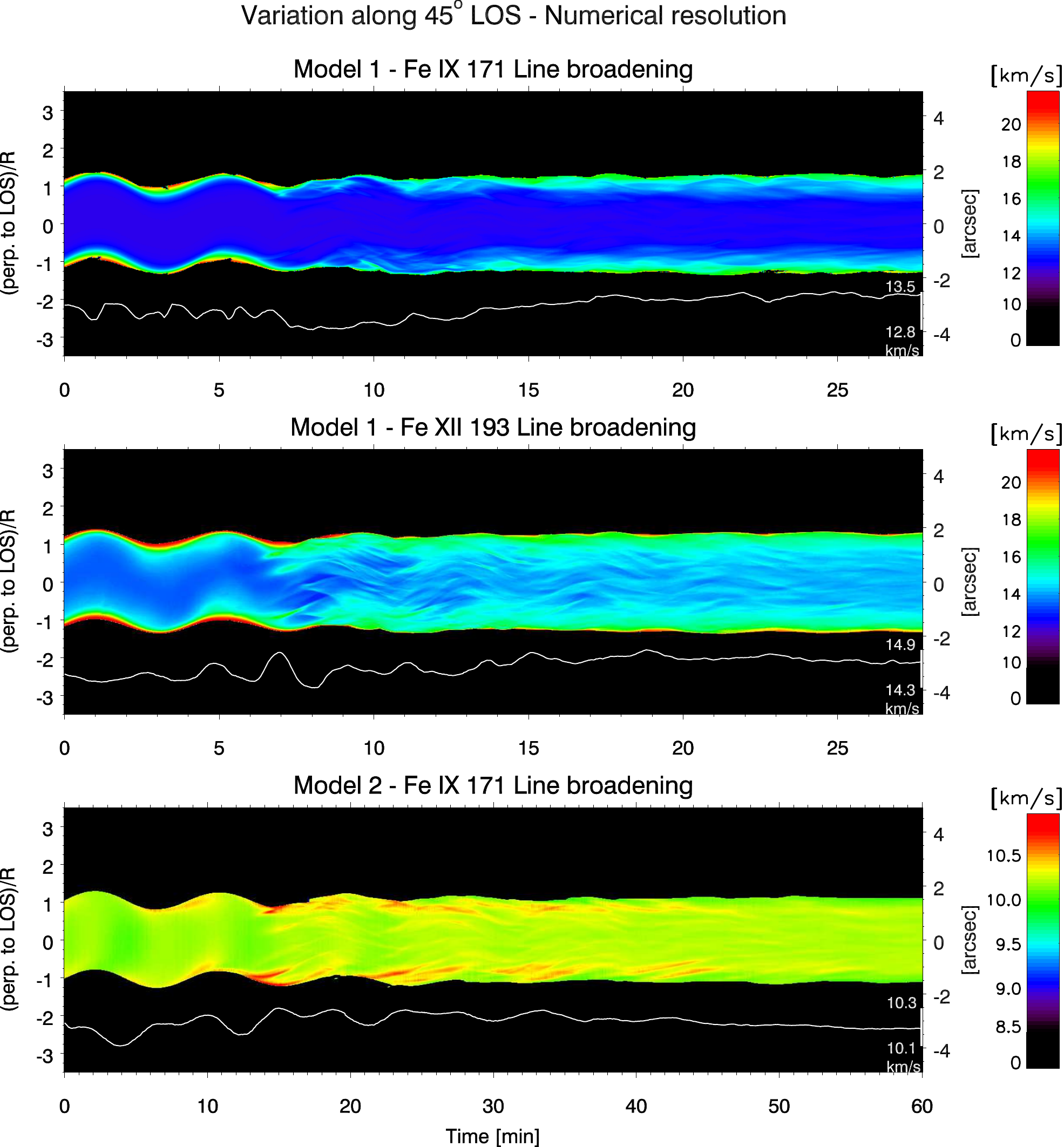}
\end{array}$
\caption{Similar to Fig.~\ref{fig9} but for the line width instead of Doppler velocity. Note that the colour scale is shifted artificially towards higher values so that the range $0-8.5~$km~s$^{-1}$ corresponds to black. The overlaid white curve corresponds to the average line width over non-zero values at each time step, smoothed over 14 sec in order to bring out the main trend. The scale and minimum and maximum values of this average are written on the lower right hand side of each panel.
\label{fig10}}
\end{center}
\end{figure}

Line broadening (including both the thermal and non-thermal components) is also relevant to detect the signatures of phase mixing (and resonant absorption), the formation of vortices and the changes in temperature. When observing at high resolution, line broadening due to the higher external temperature appears at the start in \textit{model~1} for both the boundary and core lines as a significant increase of 5~km~s$^{-1}$ from the loop core (around $13-14$~km~s$^{-1}$) to the edges (around $18-20$~km~s$^{-1}$). This is shown in Fig.~\ref{fig10} (upper and middle panels). Accordingly, in \textit{model~2} this increase is largely absent at the beginning.

The TWIKH rolls can be clearly distinguished in the line broadening maps in both lines and models as a rapid enhancement around the loop edges, after about 1 period from the onset of the KHI. This enhancement is stronger in \textit{model~1}, in which it appears uniform over time and all periodicity seems to be lost. This enhancement is produced by an increase in temperature evidenced by a decrease (increase) in line emission in the core (boundary) line, seen in Fig.~\ref{fig5}, and also by unresolved local motions along the LOS from the vortices. In \textit{model~2}, for which the temperature is mostly uniform (and therefore only the non-thermal component is visible), the maps show a smaller range of variation, which is mostly due to the unresolved motions of the vortices and also from phase mixing.  Indeed, the overlap of vortices producing the oscillatory behaviour is a result of the overlap of azimuthal Alfv\'en waves at the boundary, which are continuously generated and in turn become unstable. Accordingly, we can see the formation of the vortices in these maps as periodic stretches of enhanced broadening of $1-2$~km~s$^{-1}$ that start at the loop edge, and extend towards the loop core. While \textit{model~1} shows an increase in the average line width in time for both lines after the onset of the KHI, \textit{model~2} shows a saturation followed by a small decrease. This difference between the models indicates that the main factor behind the long term increase of the line widths in \textit{model~1} is the thermal component rather than the non-thermal one, and is due to the mixing of plasma in the non-uniform temperature model.

In the first few oscillations, we see a periodic increase of the line broadening at the trailing edge of the loop at times of maximum displacement, more clearly visible for \textit{model~1} and for a LOS angle of $0^{\circ}$. At the same times in \textit{model~2}, we can also see a small periodic increase of about $0.5~$km~s$^{-1}$ across the loop (visible as slightly more yellow regions in the figure, due to the overall small line width variation in this model). These enhancements have double periodicity (with respect to the kink mode period) and are due to the deformation of the flux tube, which increases the column depth along the LOS and therefore the unresolved motions. After the onset of the KHI, this effect is altered due to the appearance of TWIKH rolls. The double periodicity remains but now the additional, dominating factor of the TWIKH rolls define the phase relation with respect to the intensity.  Since the TWIKH rolls appear mostly at times of maximum shear velocity but converge at the trailing edge of the loop, it makes the phase relation LOS dependent. We find that for a $0^{\circ}$ LOS the intensity and line width are initially out-of-phase by $\pi$ for both the core and boundary lines \citep[as also obtained by][]{Yuan_2016ApJS..223...23Y}, meaning that the line width has maxima at times of maximum displacement. However, after the KHI they become $\pi/2$ out-of-phase only for the boundary line, the line width peaking ahead of the intensity. This advance in the increase in line width seems to be due to the new set of oppositely directed vortices appearing already before reaching maximum displacement (see Movie~1), which are only picked out in the boundary line.

\section{Predicting observable features}\label{prediction}

\subsection{Imaging characteristics at low spatial resolution and different LOS}\label{decayless}

One of the main questions we want to address is which of the observational signatures described in the previous sections could be actually observed with current resolution instruments, as well as with the next generation of detectors. We therefore look at the effect of degrading both the spatial and spectral resolution.

Degrading the temporal resolution has little effect on the results. This is because the time-scale of most of the observed features, especially at lower spatial resolution, is on the order of half a period or 1 period (255~s and 530~s for \textit{model~1} and \textit{model~2}, respectively), a time-scale much longer than usual instrument cadences. However, the smallest scales in our model, obtained in the turbulent cascade of the TWIKH rolls, do have smaller time-scales. Their effect on the intensity and Doppler velocity is however minimal, and contributes little to the line widths. 

\begin{figure}[!ht]
\begin{center}
$\begin{array}{c}
\includegraphics[scale=0.43]{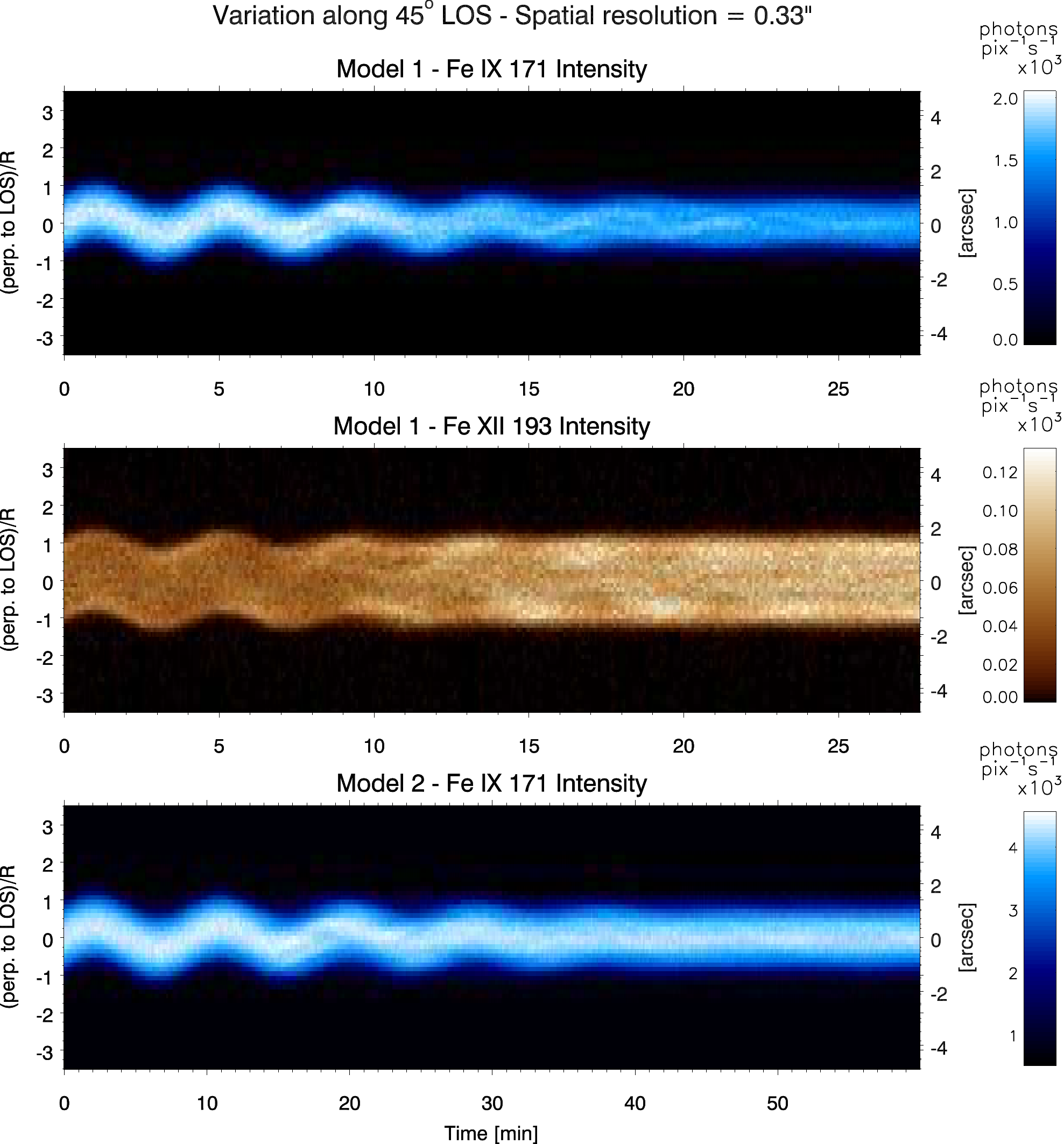}
\end{array}$
\caption{Time-distance diagrams for a slit in the same configuration as in Fig.~\ref{fig5} for the loop of \textit{model~1} in the \ion{Fe}{9} (\textit{top panels}) and \ion{Fe}{12} (\textit{bottom panels}) intensity lines, but mimicking an imaging instrument such as \textit{Hi-C} (with a spatial resolution of $0.33''$ and a cadence of 5~sec). Pixel sampling at half the resolution and 10\% photon noise are taken into account.  \textit{Model~2} seen in the \ion{Fe}{9} emission line and with the background emission subtracted. 
\label{fig11}}
\end{center}
\end{figure}

\begin{figure}[!ht]
\begin{center}
$\begin{array}{c}
\includegraphics[scale=0.43]{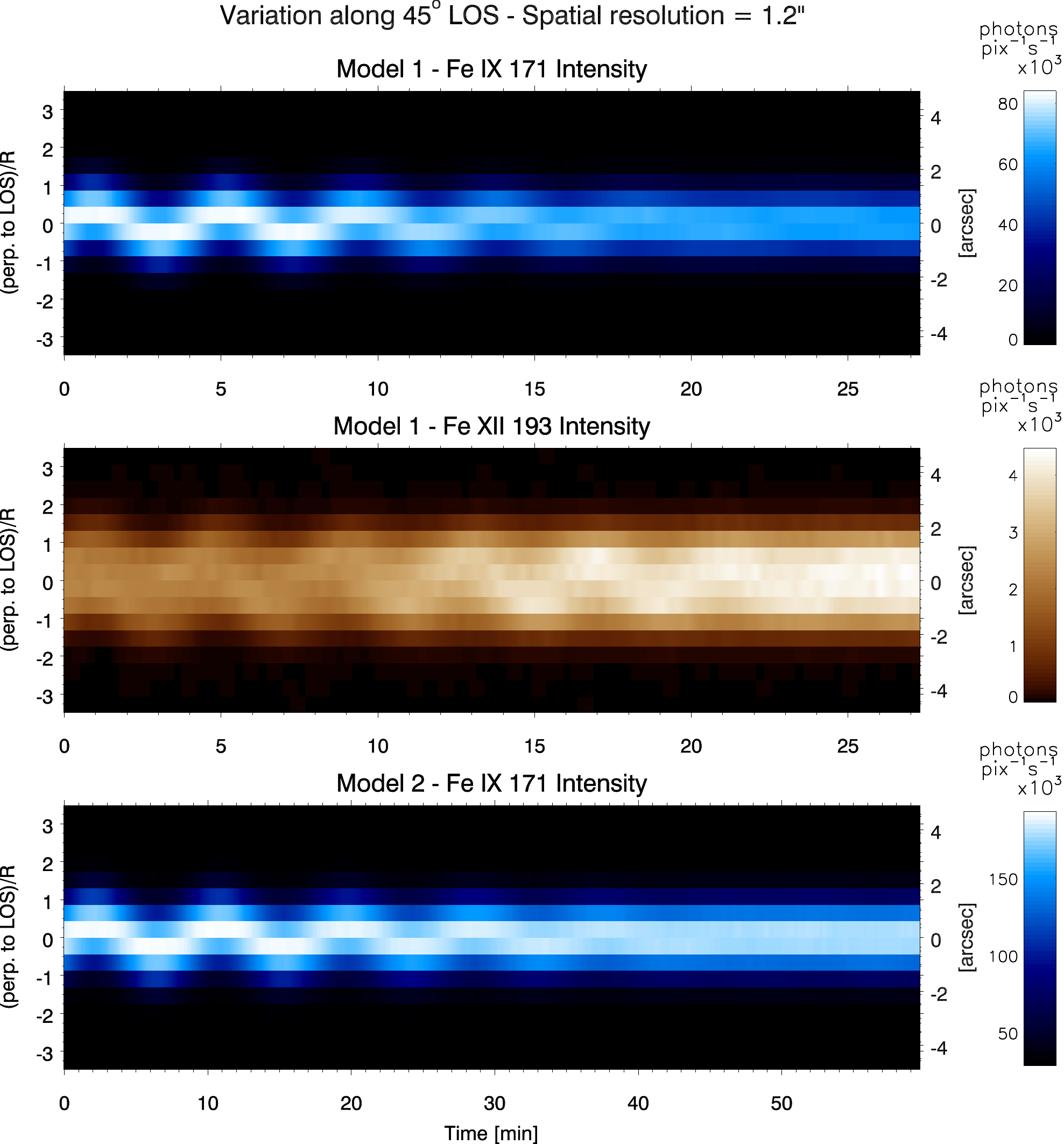}
\end{array}$
\caption{Same as Fig.~\ref{fig11} but mimicking an instrument such as \textit{SDO}/AIA (we take a spatial resolution of $1.2''$ and a cadence of 15~sec).
\label{fig12}}
\end{center}
\end{figure}

Figure~\ref{fig11} shows time-distance diagrams for both the core \ion{Fe}{9}~171 and boundary \ion{Fe}{12}~193 lines for a fixed LOS angle of 45$^{\circ}$ mimicking an instrument such as \textit{Hi-C}, with a spatial resolution of $0.33\arcsec$ and 5 second cadence. We can see that the fine loop substructure is still visible in all lines and models, as well as the overall intensity trend of thinning/broadening and fading/brightening in the core/boundary lines, and the periodic intensity changes (particularly in the boundary line) produced by resonant absorption and the KHI. We now further degrade the spatial resolution to $1.2\arcsec$ and the temporal resolution to 15~sec, representing \textit{SDO}/AIA and show the results in Fig.~\ref{fig12} for the same settings as in Fig.~\ref{fig11}. In this case, only the overall fading/brightening and the corresponding thinning/broadening is observed. The damping in the core line and decay-less oscillation in the boundary line can be clearly observed. The damping in the core line for both models looks very similar, with only a slightly longer damping time in \textit{model~2} compared to \textit{model~1} (damping time over period ratio being 2.36 and 2.30, respectively).

\begin{figure}[!ht]
\begin{center}
$\begin{array}{c}
\includegraphics[scale=0.43]{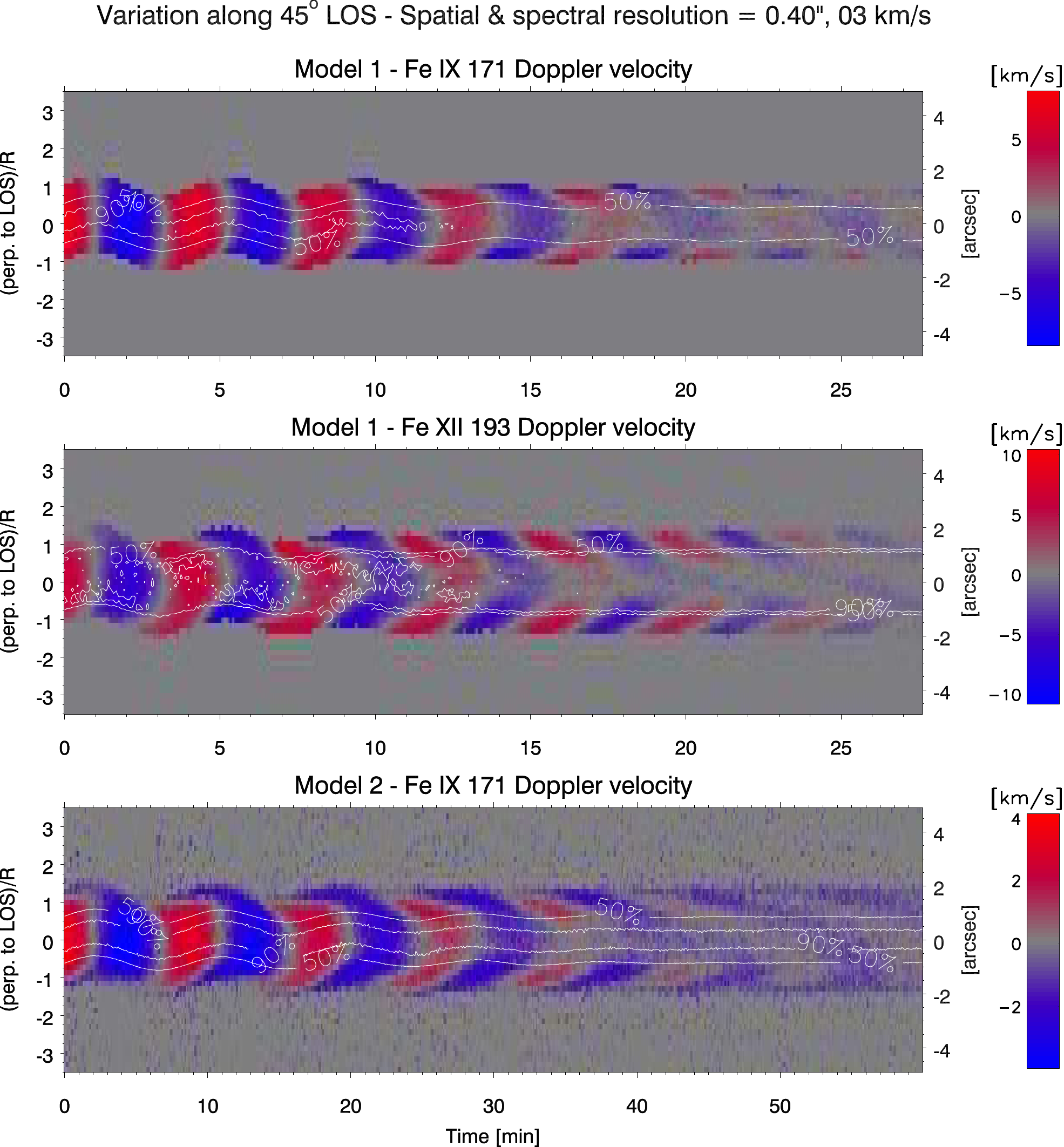}
\end{array}$
\caption{Similar to Fig.~\ref{fig11} but for the Doppler velocity. The target instrument has a slightly different spatial resolution of $0.4''$ and a spectral resolution of $3~$km~s$^{-1}$. The intensity contours with the corresponding spatial resolution are overlaid in each diagram, enclosing the regions with 50\% and 90\% intensity signal with respect to the maximum intensity at $t=0$.
\label{fig13}}
\end{center}
\end{figure}

The fact that the decay-less effect \citep[described in detail in][]{Antolin_2016ApJ...830L..22A} is basically not observed for \textit{model~2} indicates that the temperature contrast between the loop and the ambient corona combined with the mixing produced by the KHI is an essential ingredient in order to obtain this effect. In theory, if the KHI can lead to efficient heating in a more realistic model (for instance, with a more realistic turbulent spectrum at higher spatial resolution and with the expected reconnection between the vortices) then we may expect the decay-less effect to occur even with initially uniform temperature cross-sections. This is simply because in such a model the vortices would be at a higher temperature, thereby achieving the same effect as in the boundary line of \textit{model~1}.

\subsection{Spectroscopic characteristics at low spatial and spectral resolutions}\label{spectra_low}

For spectral instruments we have assumed coarser spatial resolutions of $0.4\arcsec$ and $3\arcsec$ (and keeping the temporal resolutions of 5~sec and 15~sec). The latter is representative of the \textit{Hinode}/EIS resolution, by taking a plate-scale of $1\arcsec$ but a PSF of $3\arcsec$. We also take 3 different spectral resolutions of $3~$km~s$^{-1}$, $25~$km~s$^{-1}$, and $36~$km~s$^{-1}$, the latter one being roughly that of \textit{Hinode}/EIS for the wavelength range of the core and boundary lines considered here. 

\begin{figure}[!ht]
\begin{center}
$\begin{array}{c}
\includegraphics[scale=0.43]{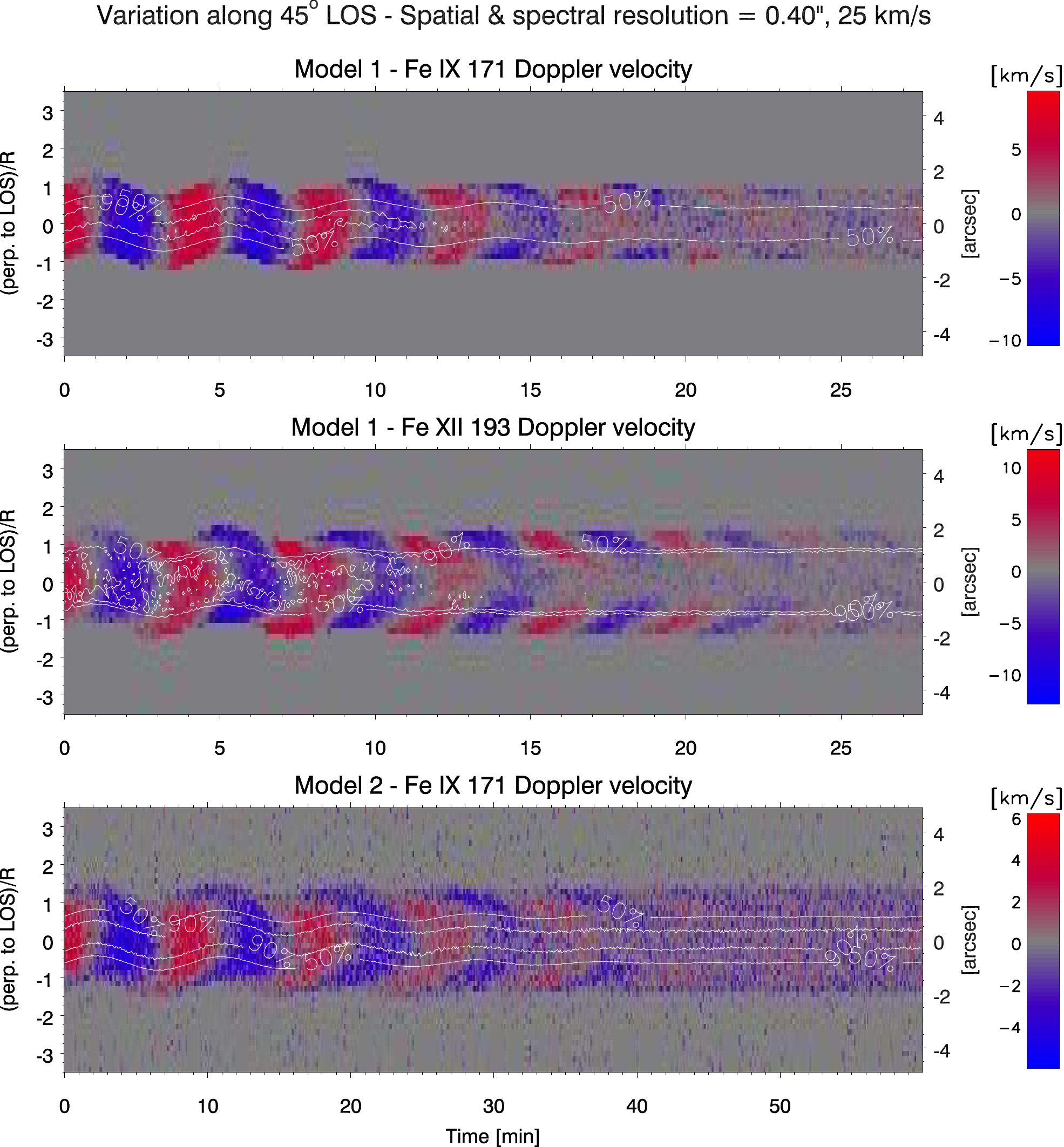}
\end{array}$
\caption{Similar to Fig.~\ref{fig15} but with a spectral resolution of $25~$km~s$^{-1}$.
\label{fig14}}
\end{center}
\end{figure}

In Figs.~\ref{fig13} and \ref{fig14} we fix the spatial resolution to $0.4\arcsec$ and take a spectral resolution of $3~$km~s$^{-1}$ and $25~$km~s$^{-1}$, respectively. In each of these figures, we have overlaid the intensity contours at various levels (the outermost being at $50\%$ of the maximum intensity at $t=0$) corresponding to the same spatial resolution (and with the other characteristics too, such as re-binning and photon noise). These contours indicate which of the spectral features could actually be observed or suffer from instrument, LOS projection and other effects. We notice that the fine structure in the Doppler signal (the ragged transition and single features from the vortices) mostly disappears, except for the case of the boundary line and only for time periods after the KHI onset. The arrow shaped Doppler structure can also be seen in all lines and models, at both spectral resolutions. However, in the core line of \textit{model~1} the arrow features are only detectable after the boundary has broadened due to the KHI. At 25~km~s$^{-1}$ spectral resolution, the Doppler transitions and arrow-shaped features fade out rapidly, and cannot be properly detected after 5 periods. The strongest Doppler signatures produced at the edge of the flux tube from resonant absorption are visible in the core line of \textit{model~1} only if the noise levels are less than roughly $15\%$ of the initial maximum intensity. In the boundary line of \textit{model~1} or core line of \textit{model~2} this condition is relaxed, and the increase of amplitude can be detected even at pixels with noise levels as much as 50~\% the initial intensity. For the boundary line this feature can be detected even at pixels with 90~\% the initial intensity. 

\begin{figure}[!ht]
\begin{center}
$\begin{array}{c}
\includegraphics[scale=0.43]{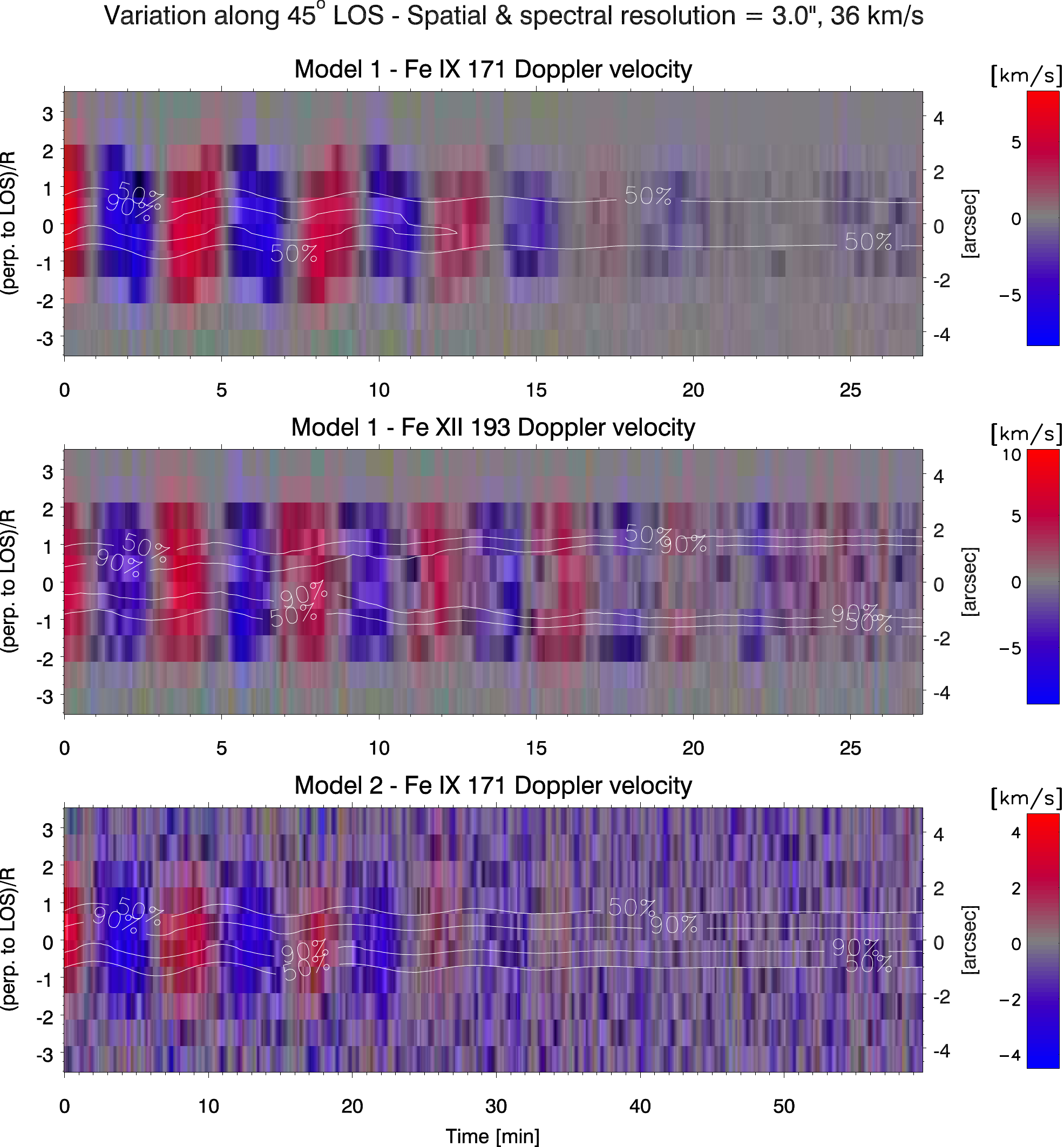}\end{array}$
\caption{Similar to Fig.~\ref{fig15} but with a spatial resolution of $3\arcsec$, a spectral resolution of $36~$km~s$^{-1}$ and a temporal resolution of 15~sec, mimicking \textit{Hinode}/EIS.
\label{fig15}}
\end{center}
\end{figure}

Figure~\ref{fig15} shows the Doppler signatures for an instrument like \textit{Hinode}/EIS. In this case, the fine-scale structure completely disappears, and the arrow-shaped Doppler profiles are only marginally visible in the boundary line at pixels with 50~\% the initial intensity. After 4 periods the periodic Doppler change becomes hard to detect in all lines and models.    

The combination of the intensity dimming/enhancement in the core/boundary lines with the arrow shape from phase mixing results in the Doppler velocity in the boundary line phasing out with respect to the Doppler velocity in the core line. This out-of-phase behaviour is still detectable in coarse resolution instruments (see Figs.~\ref{fig13},\ref{fig14},\ref{fig15}).

It is important to notice that at low spatial resolution, the maximum observable Doppler shifts at the boundary, which should actually be on the order of the initial amplitude of the kink mode (but can become significantly larger at  small scales due to the resonance), are a factor of 2 smaller than the initial perturbation. This is important when we try to reconstruct the initial kink mode perturbation from observations, and may explain the relatively small Doppler shifts from kink waves in non-flaring events, detected with \textit{Hinode}/EIS \citep{VanDoorsselaere_etal_2008AA...487L..17V}. On the other hand, we obtain roughly the same Doppler range in Figs.~\ref{fig13}, \ref{fig14} and \ref{fig15}, indicating that lower spectral resolution does not have a significant effect on capturing the main features in the Doppler signals. This is because even with a low number of points across the spectrum the line centre can still be calculated fairly accurately from Gaussian fitting and centroiding. 

\begin{figure}[!ht]
\begin{center}
$\begin{array}{c}
\includegraphics[scale=0.43]{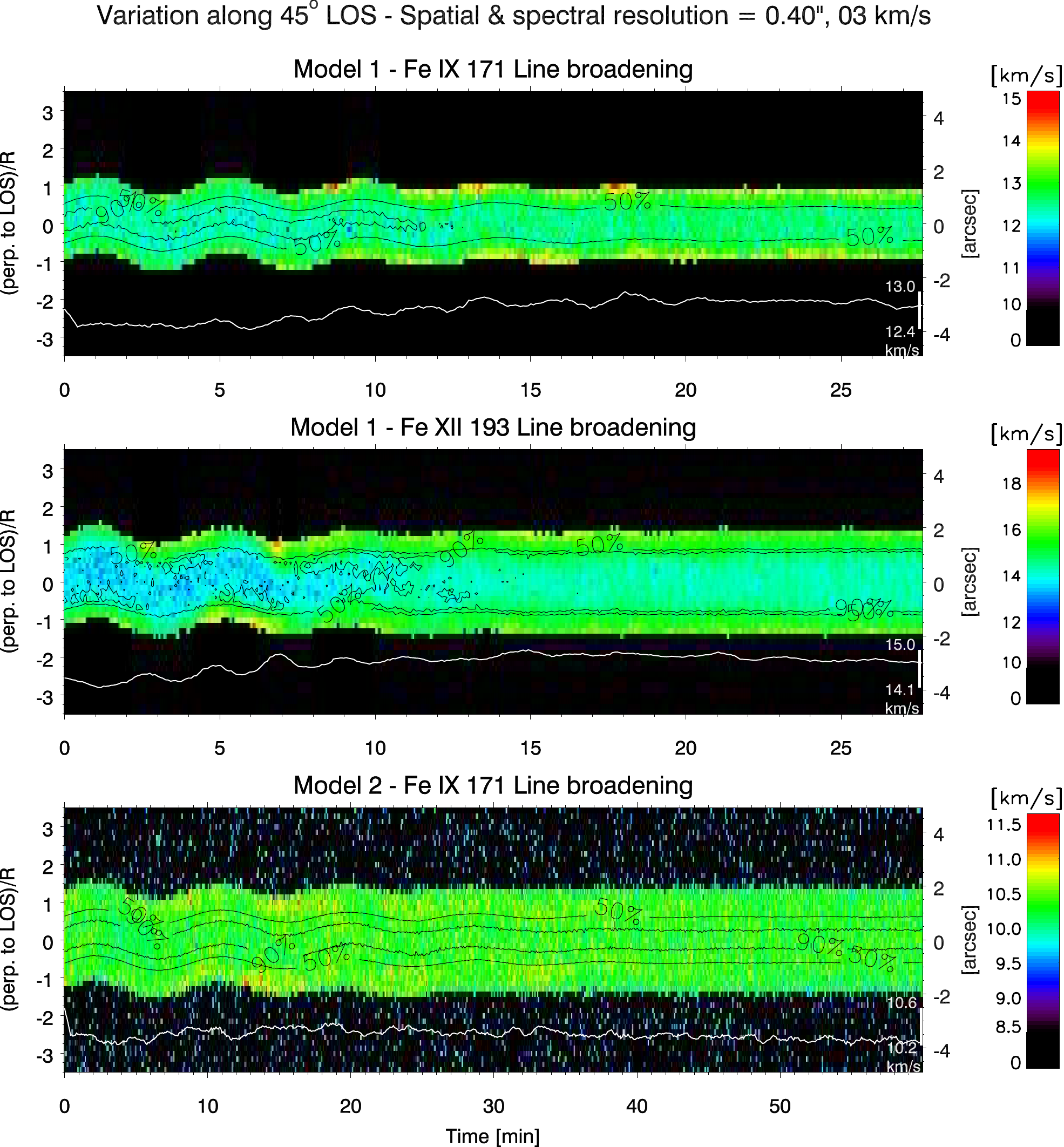}
\end{array}$
\caption{Similar to Fig.~\ref{fig13} but for the line width (see also Fig.~\ref{fig10} for further explanation). The target instrument has a spatial resolution: $0.4\arcsec$ and a spectral resolution of $3~$km~s$^{-1}$. The intensity contours with the corresponding spatial resolution are overlaid in each diagram, enclosing the regions with 50\% and 90\% intensity signal with respect to the maximum intensity at $t=0$.
\label{fig16}}
\end{center}
\end{figure}

\begin{figure}[!ht]
\begin{center}
$\begin{array}{c}
\includegraphics[scale=0.43]{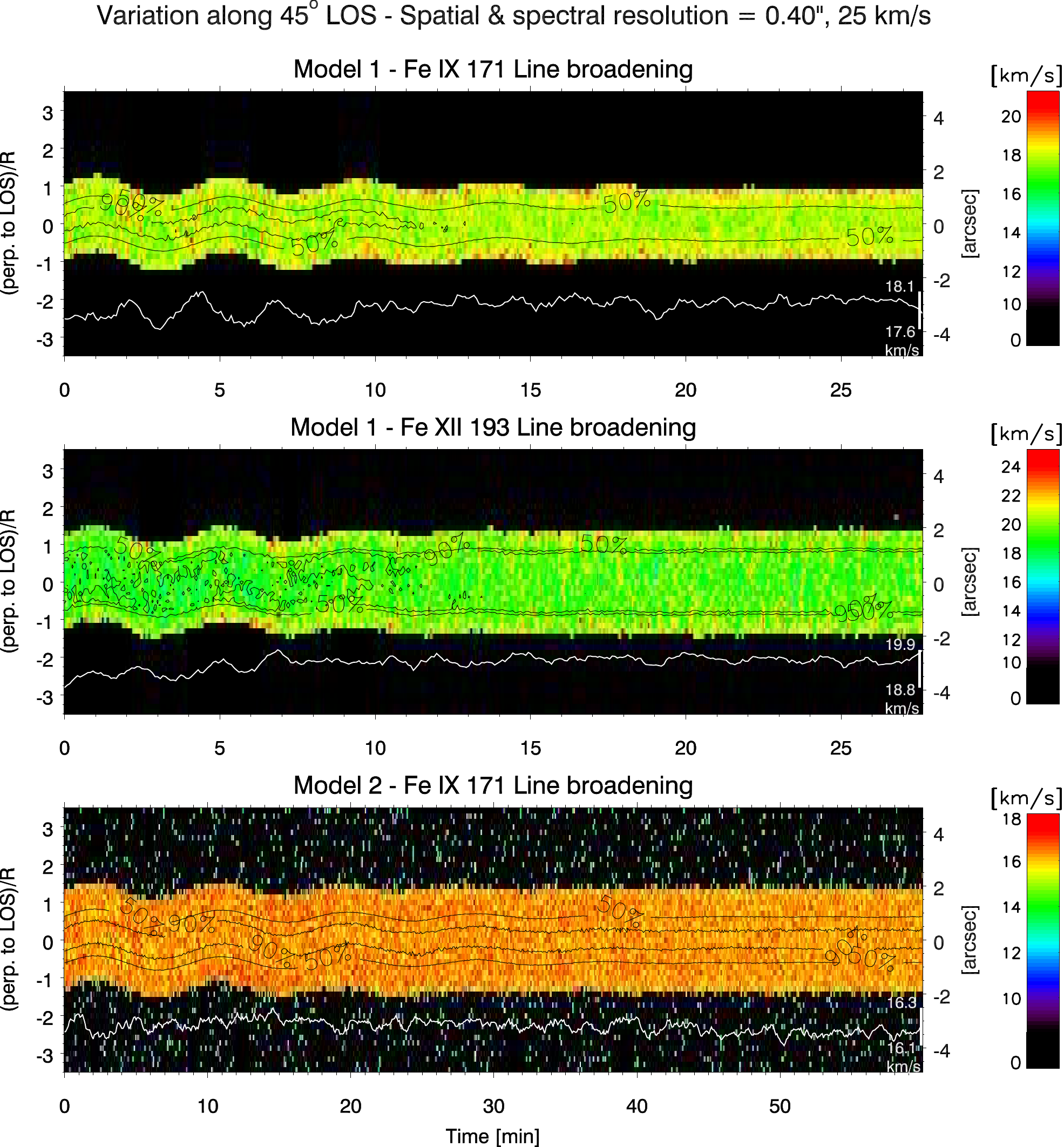}\end{array}$
\caption{Similar to Fig.~\ref{fig17} but with a spectral resolution of $25~$km~s$^{-1}$.
\label{fig17}}
\end{center}
\end{figure}

\begin{figure}[!ht]
\begin{center}
$\begin{array}{c}
\includegraphics[scale=0.43]{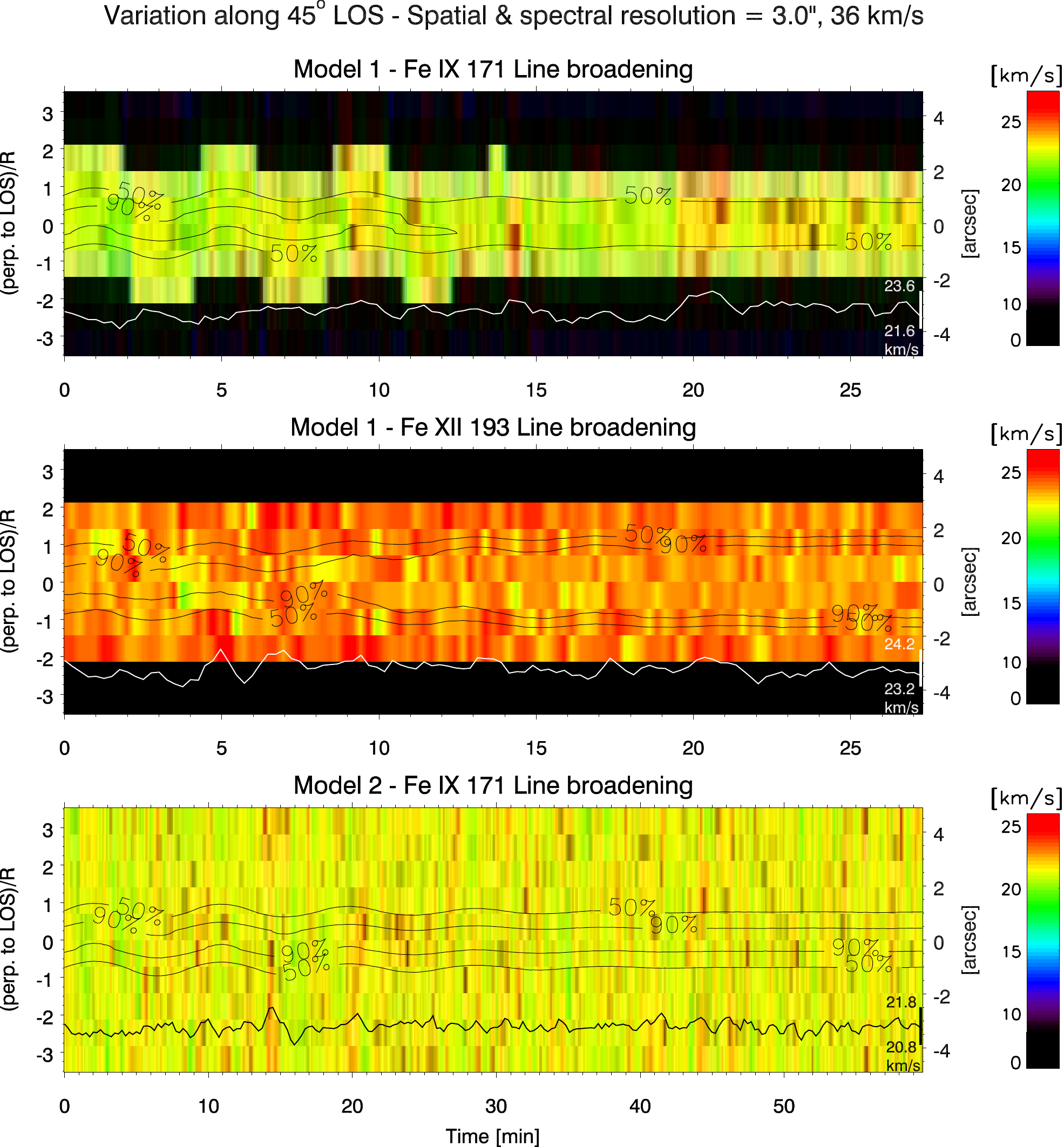}\end{array}$
\caption{Similar to Fig.~\ref{fig17} but with a spatial resolution of $3\arcsec$, a spectral resolution of $36~$km~s$^{-1}$, and a temporal resolution of 15~sec, targeting \textit{Hinode}/EIS.
\label{fig18}}
\end{center}
\end{figure}

Figures~\ref{fig16}, \ref{fig17} and \ref{fig18} show the corresponding line width measurements with the same settings as for the Doppler figures, \ref{fig13}, \ref{fig14} and \ref{fig15}, respectively. At high spectral resolution, the line broadening at the loop edges can be seen in all lines and models, but particularly in \textit{model~1}. The $1-2$~km~s$^{-1}$ enhancement from the non-thermal component is still visible in \textit{model~2}, but its periodic variation is only visible for about 2 periods after KHI onset. The change from pre to post-KHI is the most visible feature in \textit{model~1}. At low spectral resolution, the line broadening enhancement at the edges in \textit{model~1} is still visible. However, the non-thermal component in \textit{model~2} is basically undetectable, indicating that the detected line broadening in \textit{model~1} is only from the combined thermal (from mixing) and non-thermal components. Even in \textit{model~1} the pre to post-KHI change in line width is less clear at low spectral resolution.

At the low spatial and spectral resolution of \textit{Hinode}/EIS, the pre to post-KHI increase of line width can still be detected, but only in the core line of \textit{model~1}. Hence, the detectability of the line width between core and boundary appears better with the lower spatial resolution of $3\arcsec$, rather than $0.4\arcsec$. This is because the larger pixel size at the boundary captures a wider variety of unresolved motions, which are then better picked up at high spectral resolution (especially when such unresolved plasma has lower intensity than the main component of the loop core). This may also explain why we cannot see the line width increase in the boundary line at low resolution, since the addition of plasma emission from the loop core (which is dynamically very different, thereby generating broader line widths) is much reduced in that line, compared to the core line. Furthermore, due to this effect but particularly because of the lower spectral resolution, an un-realistic increase of line widths in all lines and models is obtained, especially for \textit{model~2}. When changing the spectral resolution from 3 to 25~km~s$^{-1}$ an almost double increase in the line widths is obtained and the line widths measured in the synthesised EIS spectra are roughly twice the true values, measured in Fig.~\ref{fig10}.  

\section{Discussion}\label{discussion}

In our non-uniform temperature case of \textit{model~1} the loop is colder than the ambient corona. It is important to note that this model differs minimally from a model in which the loop is both hotter and denser than the ambient corona (but which presents a corresponding lower magnetic field strength in order to maintain pressure balance). Indeed, both the resonant properties and KHI threshold of the loop remain largely the same. By setting the ambient temperature to $1$~MK in such a model, and the internal loop temperature to $1.5$~MK very similar forward modelling results would then be obtained with the same pair of emission lines, except that the results corresponding here to the core and boundary lines would be switched. 

A series of physical properties have been identified leading to loop width, intensity, Doppler and line width variations and specific phase relations, all summarised in Table~\ref{table1}. In the present model, which includes non-linear effects, we have shown that the main factors are the combination of the Kelvin-Helmholtz instability and resonant absorption (at least for the amplitudes considered here which match currently observed, non-flare related amplitudes), and to a minor degree the inertia and fluting modes produced by the initial kink mode. As shown in \citet{Antolin_2015ApJ...809...72A} the TWIKH rolls carry over characteristics of the resonant absorption and phase-mixing mechanisms, allowing these to be detected at large, observable scales. 

We find that the intensity variation in the core and boundary lines in \textit{model~1} are mostly due to the mixing between the interior and exterior plasma, which are at different temperatures. Indeed, the results from the uniform temperature model suggest that there is very little actual wave dissipation produced by the KHI in the current model, at least at the loop apex where the TWIKH rolls are bigger (and where therefore viscous dissipation is stronger) and where this study has focused. It remains an open question whether at higher resolution and different resistivity (for instance, with anomalous resistivity) efficient wave dissipation and overall heating is achieved \citep{Howson_etal_2017,Karampelas_2017}. The growth of the unstable modes is not the same along the loop and the generated vortices are larger in amplitude close to the apex, and smaller toward the footpoints. The difference in amplitude with height produces a local twist in the loop at the place of the roll-up. As the vortices degenerate into smaller and smaller structures, so does the local twist, reflected in multiple current sheets where neighbouring magnetic field lines have a small but non-zero angle with each other. We hypothesise that component magnetic reconnection could occur in this configuration, leading to possible complex braiding on scales set by the TWIKH rolls, and to efficient heating. 

In \citet{Antolin_2016ApJ...830L..22A} we have shown that at a spatial resolution corresponding to AIA, the detected damping can vary depending on the emission line for loops with non-uniform temperature cross-section, which can lead to an out-of-phase behaviour between the boundary and core lines. For a model with a cold core, hotter emission lines will observe shorter periods. This effect is produced by phase mixing and the temperature gradient in the boundary, which allows a temperature specific channel (such as that of AIA 171) to pick-up the azimuthal Alfv\'en waves having a particular period in the boundary layer corresponding to that temperature. Such temperature dependent difference in phase can therefore be used to obtain an estimate of the local density in the boundary layer for which the emission line is strongest. MHD seismology could be performed to investigate the density inhomogeneity in coronal loops with non-uniform temperature cross-section. Correspondingly, here we have shown that the same effect leads to an out-of-phase behaviour between the Doppler velocities of both lines. Also, the wavelet analysis of the intensity modulation indicates a broad power peak at the kink period compared to a thin, stronger power peak at half that period (produced mainly by the KHI) only for the non-uniform temperature model. Since phase mixing depends only on the density gradient and is thus the same in both models, this effect is purely due to temperature. In the uniform temperature model the 171 line picks up the intensity changes produced by the TWIKH rolls over a broader region encompassing both the core and the boundary, therefore leading to a broader power peak at half the kink period. Wavelet analysis could therefore be used as a diagnostic of temperature variation in the loop cross-section. 

For spectral instruments, the Doppler variation and line width shown by both the core and boundary lines in the non-uniform temperature model also present important differences linked to the main physical mechanisms. A characteristic arrow shaped structure in time-distance diagrams is observed due to the combined effects of the KHI, resonant absorption and phase-mixing. Resonant absorption strongly damps the Doppler shifts of the global kink mode (which is barely visible after 3 periods or so), transferring most of the power to the azimuthal Alfv\'en waves in the boundary layer. At the same time, the KHI broadens the boundary layer, the vortices carry the momentum of the resonant flow, producing a ragged Doppler transition (fine-scale structure) at high spatial resolution, and a maximum Doppler signal towards the edges of the flux tube which can dominate the overall signal. Phase mixing leads to the characteristic arrow-shaped structure, meaning that the Doppler velocity will show $\pi/2-\pi$ phase difference with the transverse (plane-of-the-sky) displacement of the kink mode. This effect can be captured for a spectral resolution of 25~km~s$^{-1}$ and spatial resolution of $0.33\arcsec$, and is mostly lost for a coarse spatial (and spectral) resolution such as that of \textit{Hinode}/EIS.

Lowering the spatial resolution to $0.4\arcsec$ significantly reduces the maximum observed Doppler velocity by roughly $30-50\%$. This is partly due to the localisation of the large Doppler values to the boundary layer, from resonant absorption, but also because of phase mixing, leading to the superposition of various layers of TWIKH rolls along the LOS. While the boundary layer expands due to KHI, thereby reducing the first effect, phase mixing increases in time, leading to only Doppler velocities of a few km/s by the end of the simulation. However, the results of the uniform model indicate that most of the wave energy is not dissipated and therefore remains in plasma motions and in the magnetic field. The estimation of the wave energy based on the observed Doppler velocities at the end of the simulation would therefore provide a very inaccurate value of about $10~\%$ the initial wave energy. It is interesting to note that a similar value is found in \cite{DeMoortel_Pascoe_2012ApJ...746...31D}, where the underestimation is due to mode coupling and LOS superposition of multiple flux tubes.

\begin{figure}[!ht]
\begin{center}
$\begin{array}{c}
\includegraphics[scale=0.42]{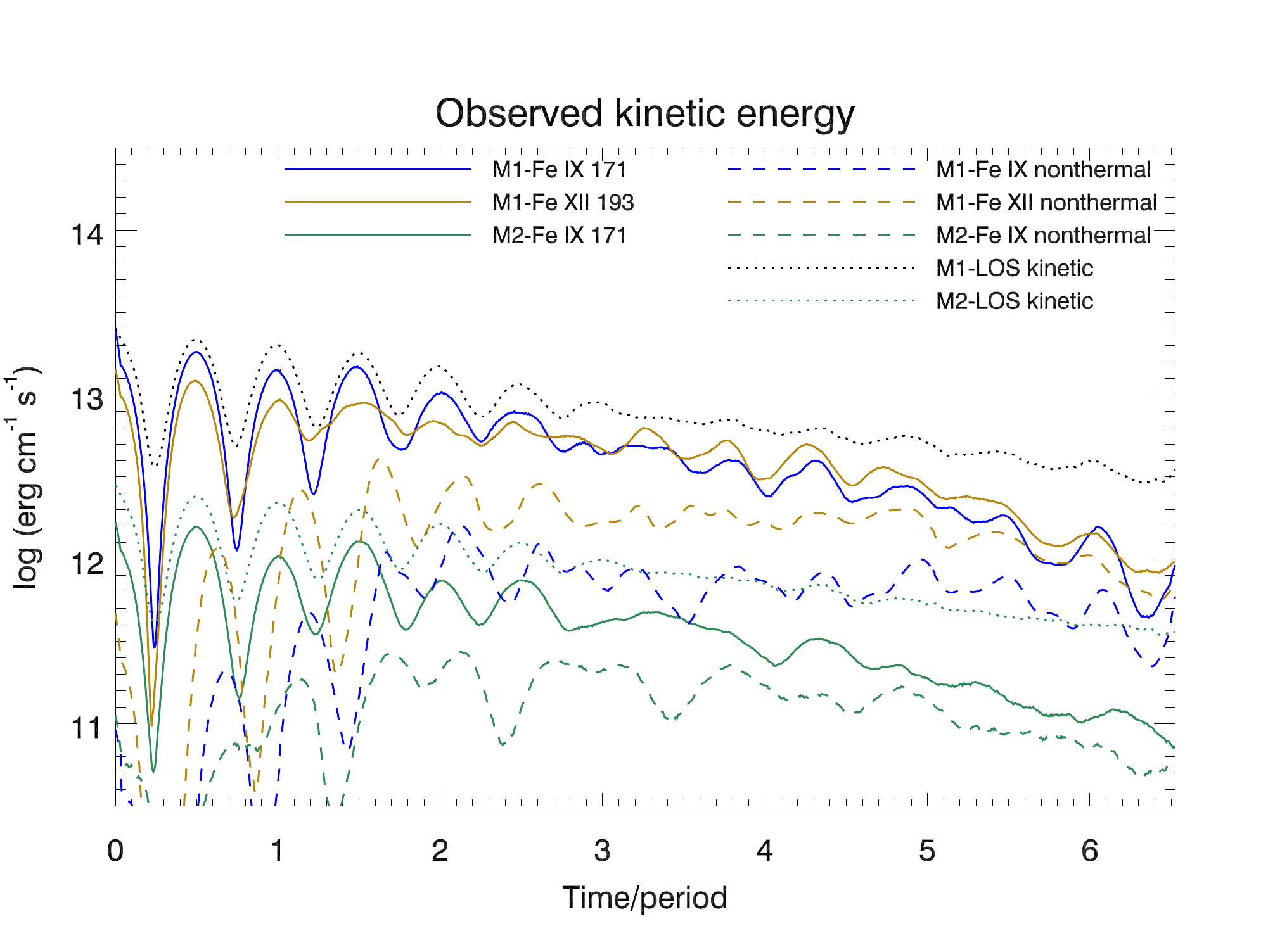}\end{array}$
\caption{The observed kinetic energy flux at the loop apex for each model (solid curves), given by Eq.~\ref{kinobs}. The true input value at the loop apex, at full numerical resolution and for a $45^{\circ}$ LOS, given by Eq.~\ref{kinput}, is shown in dotted lines for each model. The dashed curves show the kinetic energy flux provided by the non-thermal component only (which is Eq.~\ref{kinobs} without the Doppler term). The blue and dark golden colours correspond to the values derived, respectively, from the \ion{Fe}{9}~171 and \ion{Fe}{12}~193 lines in \textit{model~1}. The green coloured curves correspond to the values derived from the \ion{Fe}{9}~171 line in \textit{model~2}.
\label{fig19}}
\end{center}
\end{figure}

As explained in the introduction, an important question is how the true wave energy content of the solar corona can be determined observationally. Figure~\ref{fig19} shows the kinetic energy in each of our models, based on the observed Doppler velocities and non-thermal line widths, for the $45^{\circ}$ LOS and at full numerical resolution. The observed energy flux in a given spectral line integrated across the flux tube at the apex is calculated as follows:
\begin{equation}\label{kinobs}
F_{obs}(t) = \frac{1}{2}\int_{\perp LOS} <\rho_{\lambda}(l,t)> \left( v_{Dop,\lambda}(l,t)^{2}+\xi_{\lambda}(l,t)^2-\xi_{\lambda,th}(l,t)^2\right)c_k dl_{\perp},
\end{equation}
where the integration is over the perpendicular direction to the LOS, $c_k$ is the kink speed, $v_{Dop,\lambda}(l,t)$ is the Doppler velocity along a LOS ray $l$ and at a given time $t$, $<\rho_{\lambda}(l,t)>$ is the emissivity weighted average of the density along a LOS ray $l$ and at a given time $t$:
\begin{equation}
<\rho_{\lambda}(l,t)>=\frac{\int_{l} \rho(l^{\prime},t) \epsilon (l^{\prime},t) dl^{\prime}}{\int_{l} \epsilon(l^{\prime},t)}dl^{\prime},
\end{equation}
where $\epsilon(l,t)$ is the emissivity along the LOS ray $l$ and at a given time $t$. Also, $\xi_{\lambda}(l,t)$ and $\xi_{\lambda,th}(l,t)$ are, respectively, the total line width in km/s and the thermal component of the line width calculated with the emissivity weighted average of the temperature along the LOS ray $l$ and at a given time $t$:
\begin{equation}
<T_{\lambda} (l,t)>=\frac{\int_{l} T(l^{\prime},t) \epsilon(l^{\prime},t) dl^{\prime}}{\int_{l} \epsilon(l^{\prime},t)}dl^{\prime},
\end{equation}
\begin{equation}
\xi_{\lambda,th}(l,t) = \frac{c}{\lambda_0}\sqrt{\frac{<T_{\lambda}(l,t)> k_{B}}{\mu m_p}},
\end{equation}
where $c$ is the speed of light, $\lambda_{0}$ is the wavelength of the emitting element at rest, $k_{B}$ is the Boltzman constant, $m_{p}$ is the proton mass and $\mu_{\lambda}$ is the atomic weight (in proton masses) of the emitting element.

In Fig.~\ref{fig19} we also show the full kinetic energy flux in the same plane along the $45^{\circ}$ LOS, divided by the width of the flux tube $w \approx 2R$:

\begin{equation}\label{kinput}
F_{tot}(t)=\frac{c_k}{w}\int_x\int_y\frac{1}{2}\rho(x,y,t)\left((v_x(x,y,t)\sin(\theta))^2+(v_y(x,y,t)\cos(\theta))^2\right)dxdy.
\end{equation}

The figure shows that the observed kinetic energy of the wave in both models and lines is roughly constant on average over the first few periods, after which it decreases by a factor of 5 by the end of the simulation, following the trend of the total kinetic energy. This behaviour can be understood by the variation of the Doppler velocities and non-thermal line widths averaged across the flux tube. The former decreases due to resonant absorption, while the latter increases due to the KHI and phase mixing. As noticed previously in Fig.~\ref{fig10}, the increase of the non-thermal component of the line width saturates about halfway through the simulation, and stays roughly constant from thereon. This is because the motions from the KHI degenerate into a turbulence-like regime, from which only small-scale low amplitude perturbations are obtained. Such motions do not alter the line width by more than a few km/s, which may explain why the detection of periodicity in the line widths associated with transverse MHD waves is very scarce in the literature. By the end of the simulation, the contribution to the kinetic energy from the non-thermal component is similar or larger than the contribution from the Doppler velocity. Importantly, the average decrease in the Doppler velocity by a factor of 10 between the first and last stage of the simulation is not compensated by a similar increase in the observable non-thermal line widths.

Since the initial kink perturbations end up in localised azimuthal motions due to resonant absorption, with phase mixing and the KHI further reducing the spatial scales, possibly leading to local turbulence-like behaviour, our model supports a positive correlation between Doppler velocities and line widths, as found for instance in \citet{McIntosh_DePontieu_2012ApJ...761..138M}. However, our results also suggest that positive correlations between Doppler velocities and line widths are limited, since a significant line width increase is not obtained. This is particularly limiting for estimating the true wave energy present in the solar corona. Based on our results, we estimate the hidden energy from the wave, to be a factor of about $5-10$ of the observed value. Our results also predict an increase in the high frequency waves and small-scales accompanying these processes, perhaps matching the observations by \citet{DeMoortel_2014ApJ...782L..34D}. This is supported by numerical results from \citet{Magyar_2016ApJ...823...82M}, for a very similar  model. Power spectra of velocity fluctuations may therefore provide an additional tool for constraining the wave energy.

Our model further predicts a positive/negative correlation between line width and intensity for a boundary/core line in a non-uniform temperature model of a loop colder than the outside. The opposite correlation would be found in the case of a hot loop. On the other hand, if the loop presents no significant temperature variation then a mild negative correlation between intensity and line width should be present (assuming that the KHI does not contribute significantly to the heating, as is the case in \textit{model~2}). Anti-correlations that may match this scenario have been found with \textit{Hinode}/EIS \citep{Scott_2011ApJ...742..101S}. 

When investigating specific events of loops undergoing a single transverse perturbation, as is often the case during flares (but limiting to cases of small amplitude perturbation), our results also predict a negative correlation in time between Doppler velocities and line width, especially when observing at high spatial resolution (enough to resolve the resonant component in the Doppler maps). Although such negative correlations have been observed in active regions with \textit{Hinode}/EIS \citep{Doschek_etal_2007ApJ...667L.109D}, caution must be taken when comparing to such reports since in our work only the transverse velocity components to the loop are considered (the LOS rays are always perpendicular to the axis of the loop). While the kink perturbation does produce longitudinal flows in our model (up to a few km/s, depending on the amplitude), they are far lower than the values found in active regions, tens to hundreds of km/s \citep[see e.g.][and references therein]{Reale_2010LRSP....7....5R}.

\section{Conclusions}

We have investigated the observational signatures of transverse MHD waves by performing 3D MHD numerical simulations. We have focused on the case of standing (global) kink waves in coronal flux tubes (with radial density structuring) and considered the signatures for both imaging and spectroscopic devices. We have further taken two different coronal models, with and without a temperature variation in the perpendicular cross-section, and considered 2 spectral lines, the `core' and `boundary' lines, catching more the dynamics of the loop core and boundary, respectively. Modulation of intensity, Doppler velocities, line width and loop width are obtained, mainly due to the combination of the Kelvin-Helmholtz instability and resonant absorption, and to a minor degree to the inertia and fluting modes produced by the initial kink mode. It is important to realise therefore that non-linear effects such as those obtained in our model can lead to line widths and loop widths modulation that would be usually associated with sausage modes \citep{Antolin_VanDoorsselaere_2013AA...555A..74A}. Long, non-periodic trends include dimming/enhancement and loop width thinning/broadening in the core/boundary line, and pre- to post KHI positive jumps of intensity, Doppler and line widths. In Table~\ref{table1} we summarise the phase relations and main features of the observable quantities, combined with previous findings \citep{Antolin_2015ApJ...809...72A, Antolin_2016ApJ...830L..22A}. Despite the relatively simple model we can see that several factors combine to create a very complex set of observable features. This illustrates how hard it must be to disentangle all physical mechanisms in real observations. Our main results are the following:

\begin{itemize}
\item Fine strand-like structure increases for spectral lines catching the boundary dynamics. 

\item The apparent decay-less oscillations produced by TWIKH rolls \citep{Antolin_2016ApJ...830L..22A} are a unique feature of non-uniform temperature models.

\item The intensity power peak is dominated by the KHI, at half the kink period. The broadening of the peak in the global wavelet transform is mostly due to temperature, and increases for uniform cross-section temperature models, thereby serving as diagnostic tool of the inhomogeneous boundary layer. 

\item A small out-of-phase periodic brightening is obtained between the core and boundary line intensities with a period double that of the kink mode. The line widths and loop widths have double periodicity, and their phases with respect to the displacement depend on the LOS angle. 

\item A characteristic arrow shaped structure is obtained in Doppler velocity time-distance diagrams, with a ragged Doppler transition at high spatial resolution, and a maximum Doppler signal towards the edges of the flux tube which can dominate the overall signal. This is accompanied by an overall spectral line broadening of a few km/s.

\item Both the displacement and the Doppler velocities of the core and boundary lines become out-of-phase over time, especially at coarse spatial resolution. 

\item The above spectral features can be captured for a spectral resolution of 25~km~s$^{-1}$ and spatial resolution of $0.33\arcsec$, and most are lost for a coarse spatial (and spectral) resolution such as that of \textit{Hinode}/EIS. When reducing the spectral resolution from 3 to 25~km~s$^{-1}$ the line widths are significantly overestimated by $40-60\%$. Overall, we find that spectral resolution affects the line width measurements much more than the Doppler velocity measurements.

\item The estimation of the wave energy based on the observed Doppler velocities at the end of the simulation would therefore provide a very inaccurate value of about $10~\%$ the initial wave energy.  The non-thermal contribution to the observed kinetic energy is similar to that from the Doppler motions. The average decrease in the Doppler velocity by a factor of 10 between the first and last stage of the simulation is not observationally compensated by a similar increase in the non-thermal line widths. We estimate this discrepancy to be a factor of about $5-10$ of the observed value.
\end{itemize}
\begin{deluxetable*}{cccc}
\tablecolumns{4}
\tablecaption{Observational signatures of the kink mode with TWIKH rolls\label{table1}}
\tablehead{\colhead{Spectral} &\colhead{\multirow{2}{*}{Quantity}} & \colhead{Small scale} & \colhead{Large scale} \\
\colhead{line} &  & \multicolumn{2}{c}{\hspace{3.5cm}(Spatial \& temporal)}
}
\startdata
\multirow{6}{*}{Core} & Morphology & stranded & monolithic \\
 & Displacement $\xi_{C}$ & P;  phases out with $\xi_B$ & Gaussian damping, $\searrow$ \\ 
 & Intensity $I_C$& P/2 (0.1\%); out/ph [$\pi$] with $I_B$ & $\searrow$ ($3\%\tablenotemark{U}-20\%$) \\ 
 & Dopper velocity & P; out/ph $[\frac{\pi}{2}-\pi]$ with $\xi_C$; ragged; enhanced boundaries ($50\%$) & arrow; $\searrow$ (90\%)\\
 & Line width\tablenotemark{*} & P/2 $(3\%)$; out/ph $[\pi]$ with $I_C$; enhanced boundaries ($10\%\tablenotemark{U}-20\%$)& $\nearrow$ $(2\%\tablenotemark{U}-5\%)$ \\
 & Loop width\tablenotemark{*} & P/2 $(4\%\tablenotemark{U}-10\%)$; out/ph with $I_C$& $\searrow$ $(3\%\tablenotemark{U}-15\%)$\\
 \hline
 \multirow{6}{*}{Boundary} & Morphology & stranded & monolithic \\
 & Displacement $\xi_B$ & P; phases out with $\xi_C$ & decay-less \\ 
 & Intensity $I_B$ & P/2 (1\%); out/ph [$\pi$] with $I_C$ & $\nearrow$ (100\%) \\ 
 & Dopper velocity & P; out/ph $[\frac{\pi}{2}-\pi]$ with $\xi_B$; ragged; enhanced boundaries ($50\%$) & arrow, $\searrow$ (90\%)\\
 & Line width\tablenotemark{*} & P/2 ($3\%$); out/ph $[\frac{\pi}{2}]$ with $I_B$; enhanced boundaries $(20\%)$ & $\nearrow$ $(5\%)$\\
 & Loop width\tablenotemark{*} & P/2 $(15\%)$; in/ph with $I_B$ & $\nearrow$ $(25\%)$
\enddata
\tablenotetext{*}{Phase is LOS dependent}
\tablenotetext{U}{Values for the uniform temperature model (\textit{model~2}).}
\tablecomments{$P$ denotes the period of the kink mode. $P/2$ denotes double periodicity. Small spatial and temporal scales denote scales on the order of 1/3 of the loop radius and time changes on the order of the kink period or less. The $\nearrow$ and $\searrow$ symbols indicate increasing and decreasing amplitudes in the long term, respectively. The percentage next to the respective variation indicates its average magnitude. `out/ph' and 'in/ph' mean out-of-phase and in-phase relations, respectively. The LOS dependent phase values are for a $0^{\circ}$ LOS. All phase relations correspond to that observed after the onset of the KHI.}
\end{deluxetable*}

\acknowledgments
This research has received funding from the UK Science and Technology Facilities Council and the European Union Horizon 2020 research and innovation programme (grant agreement No. 647214), and also from JSPS KAKENHI Grant Numbers 25220703 (PI: S. Tsuneta) and 15H03640 (PI: T. Yokoyama). T.V.D. was supported by FWO Vlaanderen's Odysseus programme, GOA-2015-014 (KU Leuven) and the IAP P7/08 CHARM (Belspo). Numerical computations were carried out on Cray XC30 at the Center for Computational Astrophysics, NAOJ. This work also benefited from the ISSI - Coronal Rain (P.I. Patrick Antolin) and ISSI-BJ meetings (P.I. V. Nakariakov \& T. Van Doorsselaere).

\bibliographystyle{aasjournal}
\bibliography{ms.bbl}  

\end{document}